\documentclass[fleqn,usenatbib]{mnras}

\usepackage[T1]{fontenc}
\DeclareRobustCommand{\VAN}[3]{#2}
\let\VANthebibliography\thebibliography
\def\thebibliography{\DeclareRobustCommand{\VAN}[3]{##3}\VANthebibliography}

\usepackage{graphicx}	
\usepackage{amsmath}	
\usepackage{amssymb}	
\usepackage{appendix}
\usepackage{newtxtext,newtxmath}
\usepackage{xcolor}

\definecolor{cottoncandy}{rgb}{1.0, 0.74, 0.85}
\definecolor{darkgreen}{rgb}{0.0, 0.5, 0.0}

\title[Galaxy clusters in MilleniumTNG]{The MillenniumTNG Project: The hydrodynamical full physics simulation and a first look at its galaxy clusters}

\author[R. Pakmor et al.]{%
\parbox{0.9\textwidth}{%
R\"udiger Pakmor$^{1}$\thanks{rpakmor@mpa-garching.mpg.de},
Volker Springel$^{1}$,
Jonathan P. Coles$^{2}$,
Thomas Guillet$^{3}$,
Christoph Pfrommer$^{4}$,
Sownak Bose$^{5}$,
Monica Barrera$^{1}$,
Ana Maria Delgado$^{6}$,
Fulvio Ferlito$^{1}$,
Carlos Frenk$^{5}$,
Boryana Hadzhiyska$^{6}$,
C\'esar Hern\'andez-Aguayo$^{1,7}$,
Lars Hernquist$^{6}$,
Rahul Kannan$^{6}$,
Simon D. M. White$^{1}$
}
\\%
\\%
$^1$Max-Planck-Institut f\"{u}r Astrophysik, Karl-Schwarzschild-Str. 1, 85748, Garching, Germany\\%
$^2$Swiss National Supercomputing Centre (CSCS), Lugano, Switzerland\\%
$^3$Physics and Astronomy, University of Exeter, Exeter EX4 4QL, UK\\%
$^4$Leibniz-Institute f\"{u}r Astrophysik Potsdam (AIP), An der Sternwarte 16, 14482 Potsdam, Germany\\%
$^5$Institute for Computational Cosmology, Department of Physics, Durham University, South Road, Durham, DH1 3LE, UK\\%
$^6$Harvard-Smithsonian Center for Astrophysics, 60 Garden Street, Cambridge, MA 02138, USA\\%
$^7$Excellence Cluster ORIGINS, Boltzmannstrasse 2, 85748 Garching, Germany
}

\date{Accepted 2022 December 06. Received 2022 December 06; in original form 2022 October 18}

\pubyear{2022}

\begin{document}
\label{firstpage}
\pagerange{\pageref{firstpage}--\pageref{lastpage}}
\maketitle

\begin{abstract}
Cosmological simulations are an important theoretical pillar for understanding nonlinear structure formation in our Universe and for relating it to observations on large scales. In several papers, we introduce our MillenniumTNG (MTNG) project that provides a comprehensive set of high-resolution, large volume simulations of cosmic structure formation aiming to better understand physical processes on large scales and to help interpreting upcoming large-scale galaxy surveys. We here focus on the full physics box MTNG740 that computes a volume of $(740\,\mathrm{Mpc})^3$ with a baryonic mass resolution of $3.1\times~10^7\,\mathrm{M_\odot}$ using \textsc{arepo} with $80.6$~billion cells and the IllustrisTNG galaxy formation model. We verify that the galaxy properties produced by MTNG740 are consistent with the TNG simulations, including more recent observations. We focus on galaxy clusters and analyse cluster scaling relations and radial profiles. We show that both are broadly consistent with various observational constraints. We demonstrate that the SZ-signal on a deep lightcone is consistent with Planck limits. Finally, we compare MTNG740 clusters with galaxy clusters found in Planck and the SDSS-8 RedMaPPer richness catalogue in observational space, finding very good agreement as well. However, {\it simultaneously} matching cluster masses, richness, and Compton-$y$ requires us to assume that the SZ mass estimates for Planck clusters are underestimated by $0.2$~dex on average. Thanks to its unprecedented volume for a high-resolution hydrodynamical calculation, the MTNG740 simulation offers rich possibilities to study baryons in galaxies, galaxy clusters, and in large scale structure, and in particular their impact on  upcoming large cosmological surveys.
\end{abstract}

\begin{keywords}
galaxies: clusters -- methods: numerical -- hydrodynamics - cosmology
\end{keywords}



\section{Introduction}
Recent large cosmological hydrodynamical simulations have been very successful in reproducing realistic galaxy populations on cosmological scales. These include Illustris \citep{Vogelsberger2014Illustris,Vogelsberger2014b}, Eagle \citep{Schaye2015Eagle}, HorizonAGN \citep{HorizonAGN2016}, IllustrisTNG \citep{TNGSpringel}, Simba \citep{Dave2019Simba}, NewHorizon \citep{DuboisNEWHORIZON2021}, Thesan \citep{KannanThesan}, and Astrid \citep{Ni2022,Bird2022}. (For a recent overview of the corresponding modelling techniques, see \citet{Vogelsberger2020}.) However, among these simulation boxes even TNG300 -- as the largest of them -- is still too small to properly study large-scale structure features such as baryonic acoustic oscillations or to contain a representative sample of massive galaxy clusters. 
Larger cosmological simulations exist, e.g.~Magneticum \citep{Dolag2016} or Bahamas \citep{McCarthy2017}, but they can only afford a mass resolution that is too low to properly resolve individual galaxies, or have not reached $z=0$ yet (Astrid). Yet it is crucial to understand baryonic physics on large scales; i.e. in galaxy clusters or in the cosmic web, to anchor the baryonic physics model with galaxies and at low redshift where we have many good observational constraints.

Moreover, even larger cosmological box simulations are needed as a basis for the interpretation of upcoming enormous galaxy surveys which aim to better constrain the cosmological parameters of the Universe to the percent level (e.g.~DES, eBOSS, DESI, or Euclid). However, such simulations with sufficiently large volumes can only follow dark matter \citep{Potter2017, Angulo2022} and require significant postprocessing to model actual observables, which for the most part are either exclusively based on baryons or are at least affected by baryonic physics.

A similar problem arises when simulating galaxy clusters. Their sparsity requires large simulation volumes in order to contain a significant number of massive clusters. To simulate them at sufficient resolution, zoom simulations are thus often employed that focus most resolution elements in a small region of interest centered around a single galaxy cluster. These zoom simulations \citep{Barnes2017,Bahe2017,Cui2018} enable the study of the internal structure of galaxy clusters and galaxy cluster scaling relations without the need to follow large volumes at high resolution. It is difficult, however, to compose truly representative samples of clusters with this technique, and to properly model their cosmological foregrounds at the same time.

This study is one of the introductory papers of our new MillenniumTNG project in which we seek to make progress on pushing direct hydrodynamical simulations of galaxy formation to much larger volume than available thus far, and on linking these hydrodynamical results to still larger dark matter only simulations, thereby allowing the hydrodynamical results to be more reliably extrapolated to cosmological scales. Simultaneously, this offers a new opportunity to study large samples of hydrodynamically simulated galaxy clusters. 

To this end, our work applies the IllustrisTNG state-of-the-art cosmological galaxy formation model in an unparalleled large simulation volume  at a mass resolution that still allows us to reasonably describe the properties of individual galaxies. Our most ambitious hydrodynamic calculation is carried out in the $500\,h^{-1}\mathrm{Mpc}$ periodic box size of the seminal Millennium simulation, thus motivating the name `MillenniumTNG' we coined for the whole project. The nearly 15 times increase of the simulated volume compared to TNG300 at slightly lower mass resolution allows our hydrodynamic calculation to be more directly compared to upcoming large volume cosmological surveys, and importantly, it enables us to calibrate and improve approximate methods to predict galaxy catalogues based on dark matter only simulations. To facilitate the latter, MillenniumTNG additionally consists of a suite of dark matter only simulations computed in the same volume, with two `fixed-and-paired' versions at each resolution that make use of a variance suppression technique that effectively boosts the statistical power of the simulated volume \citep{Angulo2016}.  Furthermore, we have computed yet much larger dark matter only models that also include massive neutrinos, with up to 1.1 trillion particles in a $3000\,\mathrm{Mpc}$ box, which is meant to propel the statistical power of our predictions into the regime probed by the upcoming surveys. 

We refer to a companion paper by \citet{Aguayo2022} for full details on the simulation suite, its data products (including lightcone outputs and merger trees) and a study of basic matter and halo clustering statistics.  In the present paper, we focus on introducing the flagship hydrodynamical full physics simulation of the MillenniumTNG project, and on giving an initial characterisation of the galaxy clusters in the simulation. This offers a first glimpse at the possibilities the project offers to understand current and future cosmological observations. In \citet{Barrera2022} we present a novel version of the L-Galaxies semi-analytic model of galaxy formation and its application to the lightcone outputs of the MTNG simulations. In further companion papers, \citet{Kannan2022} analyses the  properties of very high redshift galaxies, while \citet{Bose2022} presents a galaxy clustering study based on color-selected galaxy samples. \citet{Hadzhiyska2022a,Hadzhiyska2022b} examines aspects of galaxy assembly bias, whereas \citet{Delgado2022} studies intrinsic alignments and galaxy shapes. \citet{Contreras2022} introduces an inference technique to constrain the cosmological parameters of MTNG from galaxy clustering. Finally, \citet{Ferlito2022} studies weak gravitational lensing both in the dark matter and the full physics runs.

\begin{figure*}
\includegraphics[width=0.85\textwidth]{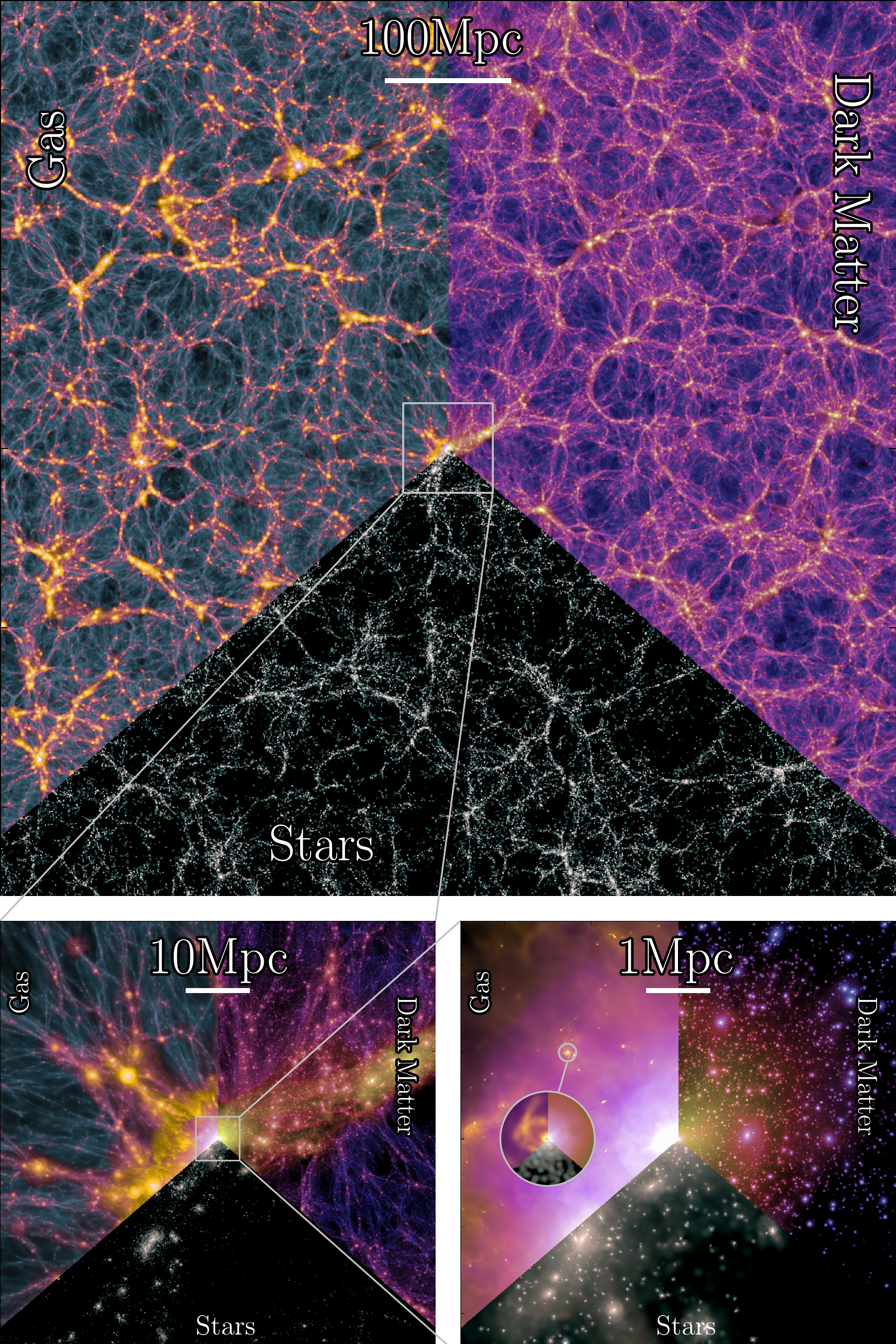}
\caption{Thin projections of gas (top left), dark matter (top right), and stellar light (bottom center) for a depth of $10\,\mathrm{Mpc}$ at $z=0$. The projections show the vast physical scales in the simulation from the full $740\,\mathrm{Mpc}$ box to an individual spiral galaxy (zoomed inset) with a radius of $50\,\mathrm{kpc}$.}
\label{fig:visuals}
\end{figure*}

This paper is structured as follows. Section~\ref{sec:simulations} introduces the MillenniumTNG full physics box and summarises its galaxy formation model and model parameters. Section~\ref{sec:verification} compares global galaxy properties of the MTNG740 full physics box to the TNG simulations, as well as updated observational data. Section~\ref{sec:scalingrelations} discusses galaxy clusters in MTNG740 and presents various galaxy cluster scaling relations and compares them to observations. Section~\ref{sec:clusterprofiles} analyses different radial profiles of galaxy clusters and puts them in relation to observations. Section~\ref{sec:lightcones} compares the SZ signal of a galaxy cluster at $z=0.25$ computed from a snapshot time-slice with spherical and cylindrical apertures to the same observable but calculated using the simulation data on a deep full backwards lightcone. Section~\ref{sec:observations} considers galaxy and cluster observables to compare the galaxy clusters in MTNG740 in observational space with clusters observed with Planck and SDSS-8. Finally, Section~\ref{sec:discussion} closes with a summary of the paper and an outlook on applications of the MTNG simulations.

\section{Full Physics Simulations in Millennium-TNG}
\label{sec:simulations}

The MTNG physics model is based on the IllustrisTNG \citep{TNGSpringel, TNGMarinacci, TNGNelson, TNGPillepich, TNGNaiman, TNG50Pillepich, TNG50Nelson} galaxy formation model \citep{TNGMethodWeinberger, TNGMethodPillepich} that has been shown to produce a fairly realistic galaxy population on cosmological scales. It includes primordial and metal line cooling \citep{Vogelsberger2013}, an explicit sub-grid model for the interstellar medium and star formation \citep{Springel2003}, mass return from stars and metal enrichment of the insterstellar medium by core-collapse supernovae, thermonuclear supernovae, and AGB~stars, an effective model for galactic winds \citep{TNGMethodPillepich}, and a model for the creation and growth of supermassive black holes as well as their feedback as active galactic nuclei \citep{TNGMethodWeinberger}.

We keep the physics model of IllustrisTNG with all its parameters unchanged, subject to only a few minor modifications. We fixed all the small issues found during and after the IllustrisTNG simulations have been run and that are documented in \citet{TNGPublicRelease}. Notably we corrected the unintentional abrupt start of the UV background radiation at the epoch of cosmic reionization. We needed to remove magnetic fields and individual metal species from the model to reduce the memory requirements of the simulation. Instead of following individual metal species we reverted to evolving only a single scalar field that tracks the total metallicity of cells and star particles. Based on small test runs, we do not expect this change to affect the simulation results for galaxies in any significant way.  Nevertheless, removing the magnetic fields does however change the model slightly \citep{TNGMethodPillepich}, and we discuss related effects in more detail in Section~\ref{sec:verification}. Since magnetic fields are omitted, we switched to using an exact Riemann solver for hydrodynamics instead of using an approximate one. In contrast to IllustrisTNG, we did not include any passive tracer particles, again for reasons of memory consumption.

Despite keeping the physics model almost identical to IllustrisTNG, we implemented a large number of technical changes and improvements to the \textsc{arepo} code \citep{Arepo, Pakmor2016, ArepoPublic} in order to make the MTNG740 full physics box fit into the memory of the supercomputer available to us. For example, for this purpose we now use shared memory on compute nodes via MPI-3 to store identical data only once per node rather than once per compute core. This includes, most importantly, information about the domain decomposition and the top-level tree that is shared between all MPI ranks, as well as various data tables such as stellar yields or photometric tables for stellar population synthesis models. We now also exploit the shared memory on nodes to improve the efficiency of various global MPI operations, for example by computing collective results first on each node, and then exchanging them only between nodes rather than between all cores individually. Similarly, we have replaced the global domain decomposition of the code with a hierarchical strategy that first subdivides the simulated volume among all nodes, and then cuts down the pieces further within each node to distribute the work among the available compute cores. Despite the use of well over $10^5$ MPI ranks, this allows us to always efficiently find a domain decomposition with a well balanced computational load and a maximum memory overhead of around $10\%$. Finally, we have reordered some operations in the gravitational tree algorithm, allowing the front-loading of a communication step. In this way, we can now almost completely avoid wait times originating from different numbers of imported and exported resolution elements on a local MPI task \citep{Arepo}.

We have included as much postprocessing as possible already on the fly during the simulation. This includes running the {\small FRIENDS-OF-FRIENDS} (FoF) group finder and a novel {\small SUBFIND-HBT} substructure finder \citep{Springel2021} adopted from the \textsc{Gadget4} into our moving-mesh code \textsc{Arepo}. It also involves  computing merger trees and matter power spectra already while the simulation is run. Finally, we output resolution elements when they intersect the past backwards lightcone of a fiducial observer, i.e. light emitted from them will be seen by the fiducial observer exactly at $z=0$. The box is periodically replicated if the lightcone extends beyond the size of the box. We use five different lightcone geometries as described in detail in \citet{Aguayo2022}.

In the following, we focus almost exlusively on the hydrodynamical full physics box of the MillenniumTNG project that we refer to as ``MTNG740'' following the naming convention of the IllustrisTNG project. This calculation follows structure formation in a cubic periodic box with side length $500\,h^{-1}\mathrm{Mpc}$ ($740\,\mathrm{Mpc}$), using $4320^3$ dark matter particles of mass  $1.7\times 10^8\,\mathrm{M_\odot}$ and initially $4320^3$ gas cells with an initial mass of $3.1\times 10^7\,\mathrm{M_\odot}$, which is the targeted baryonic mass resolution. This makes MTNG740 $15$ times larger than TNG300 by volume at a mass resolution that is only $2.8$ times worse. The minimum gravitational softening length for gas cells is set to $\epsilon_\mathrm{gas,min}=370\,\mathrm{pc}$ (it changes with the cell size), while the gravitational softening length of dark matter and stars is set to $\epsilon_\mathrm{DM,\star}=3.7\,\mathrm{kpc}$.

MTNG uses the Planck~2016 cosmology \citep{Planck2016} in identical form as IllustrisTNG in order to facilitate direct comparisons with TNG, i.e.~$\Omega_{0}=0.3089$, $\Omega_{\rmn{b}}=0.0486$, $\Omega_{\Lambda}=0.6911$, $\sigma_8 = 0.8159$, $n_s = 0.9667$, and $h=0.6774$. We generate the initial conditions at $z=63$ with second-order Lagrangian perturbation theory using \textsc{Gadget4} \citep{Springel2021}. We apply the fixed-and-paired variance suppression technique \citep{Angulo2016}, although we can only afford to run one realization of the pair, which gives however already a good part of the benefit of this approach. For the matching dark matter simulations, we have simulated both pairs, however. 

The MTNG740 simulation was executed on $122,880$ cores on the SuperMUC-NG machine at the Leibniz Computing Center, for a total wallclock time of $57$ days. It consumed $1.7\times 10^8$ core-hours and had a total memory requirement of $180\,\mathrm{TB}$. The total data output of the simulation amounts to $1.1\,\mathrm{PB}$.

In the following, we refer to the radius that encloses an average density of $500$ times the critical density of the universe as $R_{500\mathrm{c}}$, and for brevity, we shall usually drop the subscript c in the rest of the paper. We do the same for the analogous quantitities $M_{500}$, $M_{200}$, and $R_{200}$.

\section{Verification}
\label{sec:verification}

\begin{table*}
\centering
\begin{tabular}{ l | c c c c c c c c }
  \hline
  Name & BoxSize [comoving] & BoxSize [phys] & Volume & $N_\mathrm{ICs}$ & $M_\mathrm{dm}$ & $M_\mathrm{baryon}$ & $\epsilon_\mathrm{gas,min}$ & $\epsilon_\mathrm{dm,\star}$\\
  & [$h^{-1}{\rm Mpc}$] & [Mpc] & [Mpc$^3$] &  & [$\rmn{M}_\odot$] & [$\rmn{M}_\odot$] & [kpc] & [kpc]\\
  \hline
  MTNG740 & $500$ & $740$ & $4.0\times10^8$ & $2\times4320^3$ & $1.7\times10^8$ & $3.1\times10^7$ & $0.37$ & $3.7$\\
  TNG300 & $205$ & $300$ & $2.7\times10^7$ & $2\times2500^3$ & $5.9\times10^7$ & $1.1\times10^7$ & $0.37$ & $1.5$\\
  TNG300-2 & $205$ & $300$ & $2.7\times10^7$ & $2\times1250^3$ & $4.7\times10^8$ & $8.8\times10^7$ & $0.74$ & $3.0$\\
  TNG100 & $75$ & $110$ & $1.6\times10^6$ & $2\times1820^3$ & $7.5\times10^6$ & $1.4\times10^6$ & $0.19$ & $0.74$\\
    \hline
\end{tabular}
\caption{Basic numerical parameters of the MTNG740 full physics box compared to TNG100, TNG300, and TNG300-2. $N_\mathrm{ICs}$ denotes the number of resolution elements in the initial conditions that include an equal number of dark matter particles and gas cells. The MTNG hydrodynamical model follows a $15$ times larger volume than TNG300 at a mass resolution that lies between TNG300 and TNG300-2.}
\label{tab:simulations}
\end{table*}

Figure~\ref{fig:visuals} shows a visual impression of the flagship MTNG740 full physics simulation.
To verify the galaxy formation results of the MTNG740 full physics box we compare them to a set of simulations from the IllustrisTNG project. We specifically compare with TNG100, which has the resolution the galaxy formation physics model was calibrated at \citep{TNGMethodPillepich}, with TNG300, which is the largest box of IllustrisTNG and has a resolution closest to MTNG740, and with TNG300-2, which has the same box size as TNG300 but features an eight times coarser mass resolution. Thus TNG300 and TNG300-2 bracket MTNG740 in mass resolution. In Table~\ref{tab:simulations}, we list the main numerical parameters of these simulations and contrast them with the MTNG740 full physics box. Notably, MTNG740 simulates a nearly $15$ times bigger volume than TNG300 at a mass resolution that is worse by a factor of $2.8$. This is made possible by following $5.2$ times more resolution elements than TNG300.

\begin{figure*}
\includegraphics[width=0.97\textwidth]{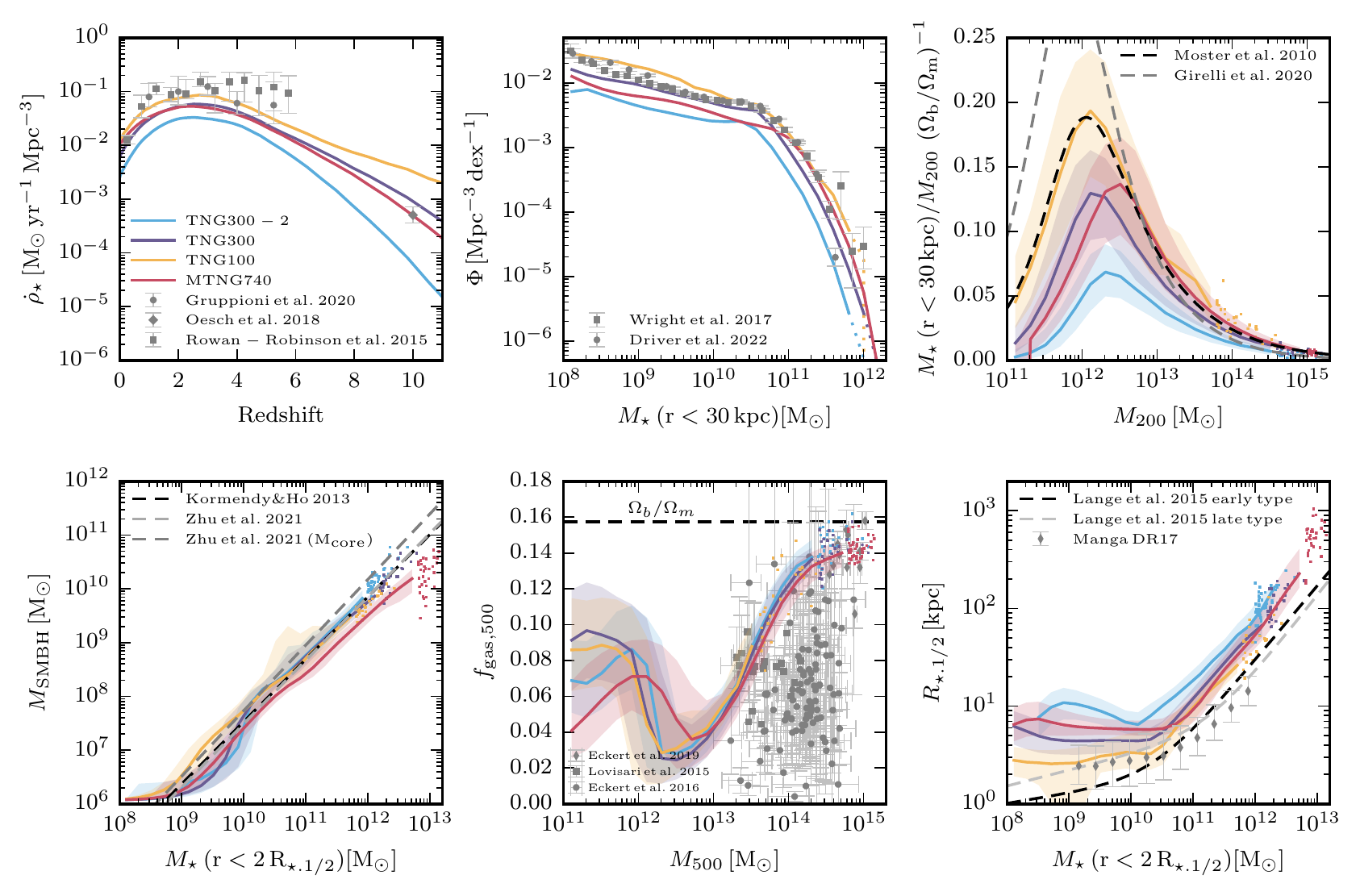}
\caption{Verification plots for MTNG740. We show the quantities used to calibrate the TNG model at the resolution of TNG100 \citep{TNGMethodPillepich} and compare to updated observational data. The individual panels show the cosmic star formation rate density (top left), the galaxy stellar mass function (top center), the stellar mass halo mass relation (top right), the mass of the central supermassive black hole at fixed host galaxy stellar mass (bottom left), the gas fraction of halos (bottom center), and the sizes of galaxies (bottom right). We compare the new MTNG740 simulation with the TNG100, TNG300, and TNG300-2 simulations at $z=0$. Galaxies are identified by {\small SUBFIND} and halos by the FoF algorithm. We show binned medians and $68\%$ percentile bands computed for four bins per decade, while individual objects are displayed instead if a bin contains fewer than ten objects. We see overall very good agreement of MTNG740 with its predecessor simulations. We also include recent observed relations and data to guide the eye, and to allow a rough qualitative assessment. In particular, at masses $M_\mathrm{200c}>10^{12.5}\,\mathrm{M_\odot}$ we see good agreement with observations, except for the hot gas fraction in all but the most massive clusters. Note that, interestingly, MTNG740 agrees best with TNG100 in the stellar mass halo mass relation, even though it is closest to TNG300 in resolution. This is related to our omission of magnetic fields in MTNG740.}
\label{fig:tngmtng}
\end{figure*}

To assess the galaxy properties computed by the MTNG740 full physics simulation, we reproduce in Figure~\ref{fig:tngmtng} the main calibration plots used  for the TNG galaxy formation model \citep{TNGMethodWeinberger, TNGMethodPillepich} for the four simulations summarised in Table~\ref{tab:simulations}. Where applicable, we bin the simulation data logarithmically with four bins per decade. When we show $16$ and $84$ percentile bands of the corresponding bins, while the solid lines show the median in each bin. We show individual objects as points rather than medians once a bin contains fewer than ten objects.

To guide the eye and to allow a first rough qualitative assessment, we also include  updated observational  data in every panel, while for  the original data used in the calibration of the model we refer to \citet{TNGMethodPillepich}. Note, however, that all data shown here are quantities inferred from observations based on additional assumptions, they are not direct observables. A more powerful comparison would require elaborate forward modelling of mock observations of the simulations, such that synthetic observables could be contrasted directly with observed quantities. This is beyond the scope of this paper but will be addressed in future work.

In the top left panel of Figure~\ref{fig:tngmtng}, we show the evolution of the cosmic star formation rate density, $\dot\rho_\star\,\mathrm{[M_\odot\,yr^{-1}\,Mpc^{-3}]}$. We compute it from the initial mass of all star particles formed in the simulations, using $100$ logarithmically spaced redshift bins. The star formation rate density of the simulations changes systematically with mass resolution such that simulations with better mass resolution form more stars. MTNG740 lies closest to TNG300, albeit slightly below it at high redshift, and slightly above it at $z=0$ where it almost matches TNG100. For comparison, we consider recent results from reconstructions of the star formation rate density of the Universe between $z=0.5$ and $z=6$ using sub-mm observations \citep{Gruppioni2020}, from infrared observations between $z=0.5$ and $z=6$ \citep{RowanRobinson2016} and at high redshift from UV data \citep{Oesch2018}. The star formation rate density in MTNG740 is slightly lower than observations for $z<6$ but in good agreement with the data points at $z=10$. Note that there are known inconsistencies between the star formation rate density reconstructed from star formation rate indicators and the total stellar mass density inferred for the Universe. The latter seems to favour a star formation rate density that is smaller by about a factor of two at $z=2$ \citep{Wilkins2008,Yu2016}.

\begin{figure*}
\includegraphics[width=0.97\textwidth]{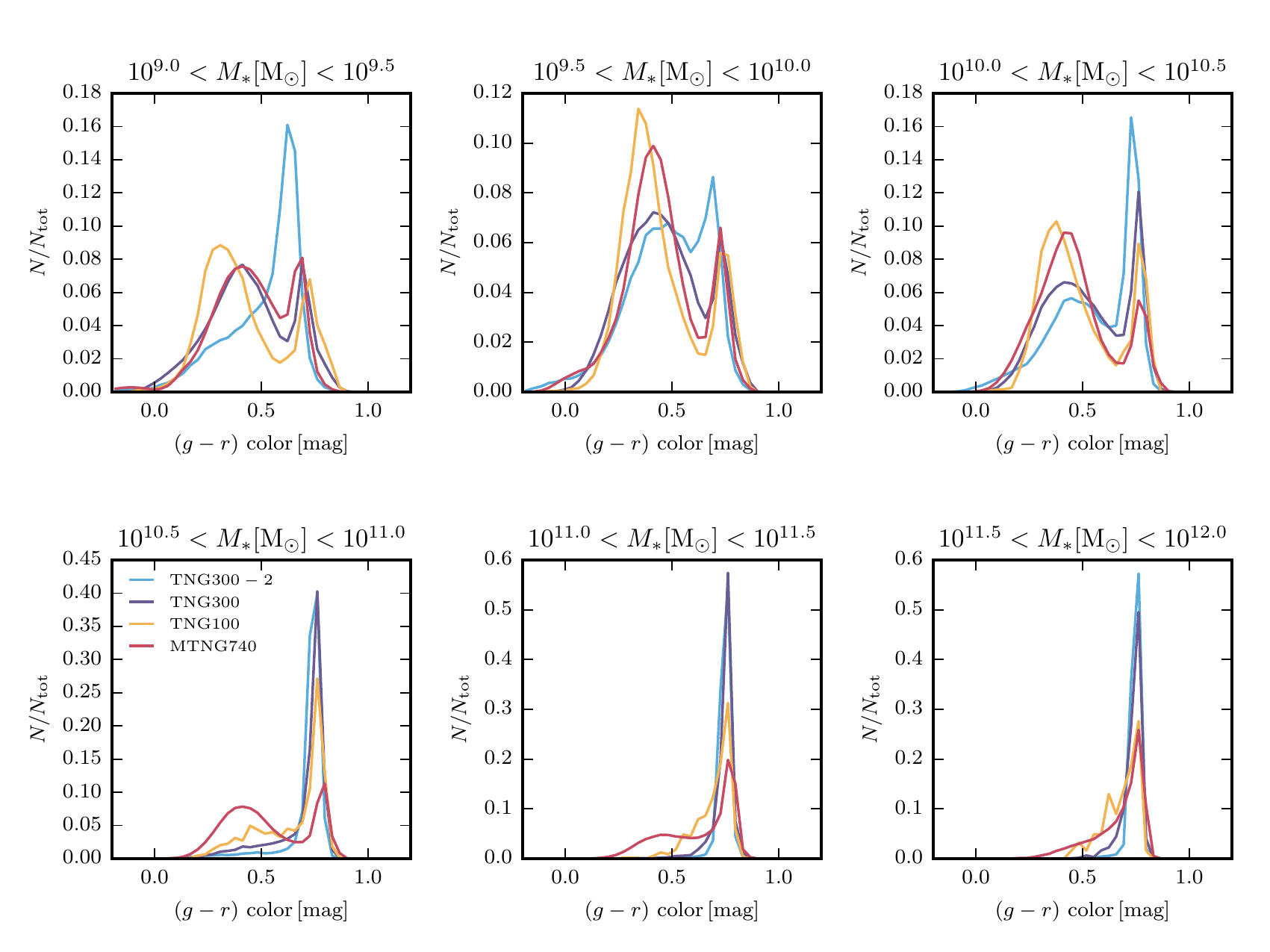}
\caption{Color distribution of galaxies in the SDSS $g-r$ color for different stellar mass bins at $z=0$. Colors are computed following \citet{Torrey2014} assuming a single stellar population for each star particle with a Chabrier IMF \citep{ChabrierIMF} and the \citet{Bruzual} stellar population synthesis model without dust. Each panel shows a normalised color histogram  of all galaxies in the corresponding mass bin, using $40$ color bins. We compare the MTNG740 simulation with the TNG100, TNG300, and TNG300-2 simulations. At the high and low mass end, MTNG740 agrees well with the TNG simulations and best with TNG100. For stellar masses between $10^{10.5}\,\mathrm{M_\odot}$ and $10^{11.5}\,\mathrm{M_\odot}$ MTNG740 has a more prominent contribution of blue galaxies compared to the TNG simulations. Thus the mass scale at which galaxies are quenched moves to slightly higher masses in MTNG740.}
\label{fig:galaxycolors}
\end{figure*}

The top center panel of Figure~\ref{fig:tngmtng} shows the galaxy stellar mass function. Here we measure the stellar mass in a fixed physical radius of $30\,\mathrm{kpc}$ that is bound to the galaxy  according to {\small SUBFIND}. Satellite galaxies are included as well. For comparison, we show the galaxy stellar mass function from the Galaxy And Mass Assembly (GAMA) survey \citep{Wright2017,Driver2022}. 

We find good agreement at the high mass end, $M_\star\gtrsim10^{11}\,\mathrm{M_\odot}$. MTNG740 is lower by about a factor of two at fixed stellar mass for galaxies with around $M_\star \sim 5\times10^{10}\,\mathrm{M_\odot}$ and about $30\%$ lower compared to observations for galaxies with stellar masses $M_\star<10^{10}\,\mathrm{M_\odot}$ .

In the top right panel of Figure~\ref{fig:tngmtng} we show the stellar mass, halo mass relation for all central galaxies and their dark matter halos. We compute the ratio of the total stellar mass within a fixed physical aperture of $30\,\mathrm{kpc}$ to $M_\mathrm{200}$ and divide by the cosmic baryon fraction. We compare the simulations to relations inferred from abundance matching that are based on galaxy counts and dark matter only simulations of a $\Lambda\mathrm{CDM}$ universe \citep{Moster2010,Girelli2020}. MTNG740 is in excellent agreement with TNG100 and abundance matching results for massive halos with $M_\mathrm{200}>10^{12.5}\,\mathrm{M_\odot}$. For less massive halos with $M_\mathrm{200}<10^{12.5}\,\mathrm{M_\odot}$, the stellar mass of galaxies in MTNG740 is smaller at fixed halo mass than for galaxies in TNG100 but comparable to galaxies in TNG300. Note that at these smaller masses there is also large variation in the estimates from abundance matching and other methods that attempt to reconstruct the relation between stellar mass and halo mass from observations \citep[see, e.g.][]{Girelli2020}.

In the bottom left panel of Figure~\ref{fig:tngmtng}, we show the relation between the mass of the most massive black hole of a galaxy and its central stellar mass. Following \citet{TNGMethodPillepich} we measure the stellar mass as the total stellar mass within twice the stellar half-mass radius, $R_{\star.1/2}$. We include central galaxies as well as satellite galaxies. We find good agreement between TNG100, TNG300, and TNG300-2. Black holes in MTNG740 are less massive by up to a factor of two compared to the TNG simulations at fixed stellar mass. For comparison, we show the observed relation between black hole mass and stellar mass of the bulge of their host galaxies that was used for calibration of the IllustrisTNG model \citep{Kormendy2013}, as well as two more recent relations between black hole mass and stellar bulge and core masses \citep{Zhu2021}. The new bulge mass relation is essentially identical to the old black hole mass bulge mass relation. The new core mass relation has the same slope but a slightly higher normalisation. We find overall reasonable agreement between MTNG740 and the observed relations. Note, however, that the stellar mass estimate we use here includes a much larger fraction (about $80\%$) of the total stellar mass of galaxies than observational estimates of bulge mass. Moreover, at the high mass end, the stellar mass estimate includes significant contributions from the intercluster light component. We discuss this in a bit more detail below in the context of the galaxy size estimates shown in the bottom right panel of Figure~\ref{fig:tngmtng}. Also note that \citet{Borrow2022} recently identified an unintended behaviour of the black hole repositioning scheme in the code that can cause massive galaxies to lose their central black holes when their gas is completely stripped by a more massive object (e.g.~within a galaxy cluster). This is also present in MTNG740,  but does not affect the median values shown here.

In the bottom center panel of Figure~\ref{fig:tngmtng}, we show the gas fraction within $R_\mathrm{500}$ and its dependence on $M_\mathrm{500}$ for halos in MTNG740, finding good agreement between the different simulations for $M_\mathrm{200}>5\times10^{12}\,\mathrm{M_\odot}$. We compare to observational data for galaxy groups \citep{Lovisari2015} and galaxy clusters \citep{Eckert2016,Eckert2019}. We find that MTNG740 is consistent with the latest data points for the most massive clusters, and also agrees on the group scale. For galaxy clusters with masses $M_\mathrm{200}\gtrsim5\times10^{13}\,\mathrm{M_\odot}$, however, the inferred gas fractions of most observed clusters are significantly below the gas fractions of the MTNG740 clusters. Moreover, the observational data has a much larger scatter at fixed halo mass than the simulations. A more detailed comparison of the gas fractions in TNG300 with observations using mock X-ray images and estimating the gas mass and total mass of galaxy clusters in the more similar way to observations also finds systematically larger values for TNG300 compared to observations, but much better agreement on the scatter at fixed halo mass \citep{Pop2022}. Additional studies will be needed to understand if the differences at $M_{500}\gtrsim10^{14}\,\mathrm{M_\odot}$ and in the associated scatter are primarily caused by observational biases, or rather reflect a limitation of the galaxy formation model in the simulations.

Finally, the bottom right panel of Figure~\ref{fig:tngmtng} we show the stellar half mass radius of galaxies depending on their stellar mass within twice the stellar half mass radius. The plot includes central as well as satellite galaxies. The stellar half mass radius is computed for all stars associated to a galaxy by \textsc{subfind}. For the central galaxies of massive halos this also includes all stars that are part of the halo but not bound to any satellite galaxy, i.e. the intracluster light. This leads to large size estimates for the most massive galaxies. The different simulations are consistent for $M_\star>10^{11}\,\mathrm{M_\odot}$, but they show a systematic trend with mass resolution for lower mass galaxies, where better resolution simulations have smaller galaxies at fixed stellar mass. For comparison, we show data from the GAMA survey \citep{Lange2015} using their radii computed from $r$-band luminosity profiles, and from $10000$ galaxies of the Manga \citep{Bundy2015} survey released in SDSS~DR17 \citep{SDSSDR17} using stellar mass and effective radius estimates from the \textsc{PIPE3D} pipeline \citep{Sanchez2016}. The simulations are reasonably consistent with the observations, although galaxies in MTNG740 appear systematically larger than the observed galaxies by about a factor of two at fixed stellar mass.

\begin{figure}
\includegraphics[width=0.97\linewidth]{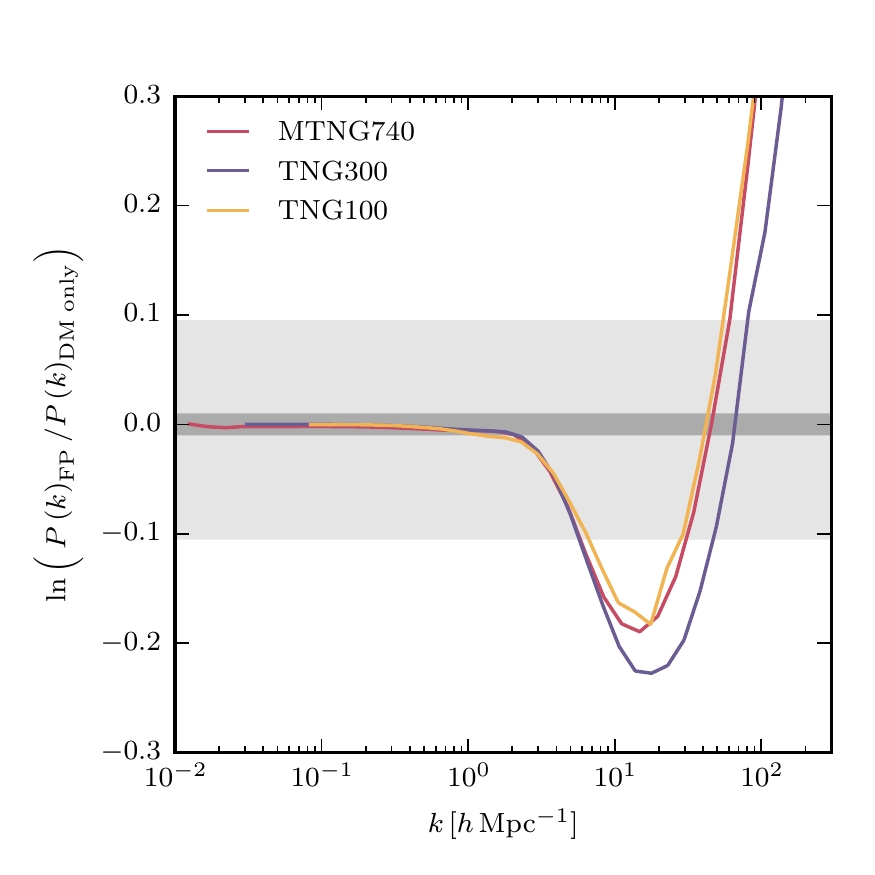}
\caption{Baryonic effects on the total matter power spectrum at $z=0$ for MTNG740, TNG100, and TNG300. For each simulation, we show the ratio of the total matter power spectrum of the full physics simulation divided by the total matter power spectrum of its dark-matter only counterpart. The grey bands show $1\%$ and $10\%$ percent deviations, respectively. MTNG740 is consistent with the TNG simulations, and is best matched by TNG100.}
\label{fig:powerspectrum}
\end{figure}

An important further diagnostic of the galaxy population is its color distribution. In particular, it is fundamental to quantify to which degree our simulated galaxy population reproduces the observed mass-dependent bimodal color distribution. Since galaxy color depends strongly on the specific star formation rate this is a good measure of quenching. TNG100 in particular has been shown to agree well with the observed color distribution \citep{TNGNelson}. In Figure~\ref{fig:galaxycolors}, we compare the distribution of $g-r$ SDSS colors of MTNG740 galaxies to the color distributions of the TNG simulations. Colors are computed assuming a single stellar
population for each star particle with a Chabrier IMF \citep{Chabrier2003} and the \citet{Bruzual} stellar population synthesis model without
dust. The luminosity of all star particles bound to a galaxy according to \textsc{subfind} is then summed up to compute the total galaxy luminosity. We find excellent agreement between MTNG740 and TNG100 for low masses $M_\star<10^{10.5}\,\mathrm{M_\odot}$ and at the highest masses, $M_\star>10^{11.5}\,\mathrm{M_\odot}$. The sharp transition that is present in all TNG simulations at a mass scale $M_\star\sim5\times10^{10.5}\,\mathrm{M_\odot}$ is however shifted to higher masses by $0.5$ dex in MTNG740, and appears washed out a bit as well. We associate this with differences in the efficiency of the kinetic AGN feedback introduced by the removal of magnetic fields from the TNG physics model \citep[see][]{TNGMethodPillepich}. The change in color distribution for galaxies with masses in the range $10^{10}\,\mathrm{M_\odot}<M_\star<10^{10.5}\,\mathrm{M_\odot}$ directly affects the color selected galaxy correlation function in this mass range, which agrees less well with observations than TNG300 \citep{Bose2022}.

As a final piece of validation of the MTNG740 model against IllustrisTNG we look at the baryonic impact on the total matter power spectrum, which is an important prediction of these non-linear simulations and can affect cosmological constraints. In Figure~\ref{fig:powerspectrum}, we show the ratio of the total matter power spectra of the full physics simulations of MTNG740, TNG100, and TNG300 divided by the total matter power spectra of their dark matter only counterparts. We find that the baryonic impact of MTNG740 is almost identical to TNG100 and slightly smaller than in TNG300, with a $1\%$ deviation at $k=2\,h\,\mathrm{Mpc}^{-1}$, a minimum just above $k=10\,h\,\mathrm{Mpc}^{-1}$ and a relative difference on this scale of $\sim20\%$.

Following \citet{vanDaalen2020} we find a suppression of the power spectrum at $z=0$ from baryonic effects at a scale of $0.5\,h\,\mathrm{Mpc}^{-1}$ of $\Delta P/P_\mathrm{DM\,only}=-0.0043$ and an average baryon fraction of all halos between $6\times10^{13} < M_{500\mathrm{c}} \mathrm{[M_\odot]} < 2\times10^{14}$ at $z=0$ of $0.86$ in MTNG740. According to the fit in \citet{vanDaalen2020}, a suppression of $\Delta P/P_\mathrm{DM\,only}=-0.0034$ for our baryon fraction would be expected, which is roughly consistent but $25\%$ smaller than our actual suppression value.

We conclude from the results in this section that there is reassuring agreement between MTNG740 and its TNG predecessors, and that most of the results found for the TNG simulations, in particular for TNG100 and TNG300, can also be expected to hold for MTNG740. At some level, one can therefore view it as a very large volume extension of the TNG suite of simulations, just as we intended.

\begin{figure*}
\includegraphics[width=0.97\textwidth]{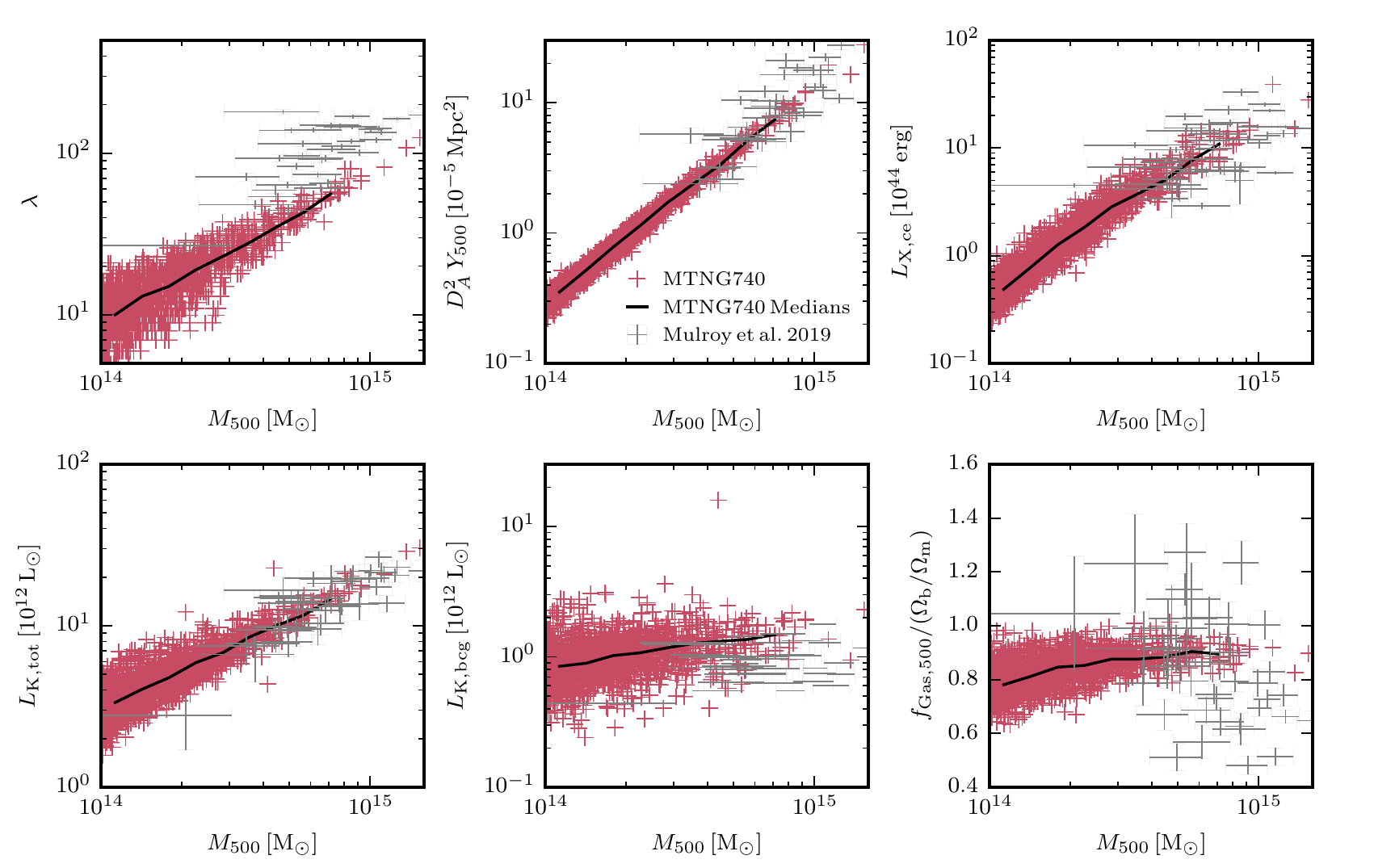}
\caption{Scaling relations for all galaxy clusters in MTNG740 with $M_{500\mathrm{c}} > 10^{14}\,\mathrm{M_\odot}$ at $z=0.25$. All quantities are computed within $R_{500}$. The panels show richness (top left), integrated Compton-$y$ parameter (top center), core excised ($0.15\,R_{500}<R<R_{500}$) bolometric X-ray luminosity (top right), total $K$-band luminosity (bottom left), $K$-band luminosity of the brightest cluster galaxy (bottom center), and gas fraction (bottom right) versus $M_{500}$. The black line shows the binned median of the MTNG740 clusters. Grey points are data of $41$ galaxy clusters between $z=0.15$ and $z=0.3$ \citep{Mulroy2019}. They are found to be in good agreement with the MTNG740 clusters.}
\label{fig:scalingrelations}
\end{figure*}

\section{Galaxy cluster scaling relations}
\label{sec:scalingrelations}

Owing to its large volume, MTNG740 contains a substantial number of massive galaxy clusters. At $z=0$ the full physics box encompasses $9$ galaxy clusters with $M_{500} > 10^{15}\,\mathrm{M_\odot}$, and more than $2000$ galaxy clusters with $M_{500} > 10^{14}\,\mathrm{M_\odot}$. At $z=0.25$, the time of the last full snapshot before $z=0$, which is important because observations cover many more clusters at this epoch than at $z=0$, the simulation already contains $3$ massive galaxy clusters with $M_{500} > 10^{15}\,\mathrm{M_\odot}$ and more than $1500$ galaxy clusters with $M_{500} > 10^{14}\,\mathrm{M_\odot}$. These numbers allow us to have a detailed look into cluster scaling relations for a representative sample of simulated galaxy clusters, and study their properties in detail.

In Figure~\ref{fig:scalingrelations}, we show six different galaxy cluster scaling relations for all galaxy clusters at $z=0.25$ with $M_\mathrm{500} > 10^{14}\,\mathrm{M_\odot}$. These relations are often used to infer the mass of a galaxy cluster from observables. We compare the galaxy clusters in MTNG740 with a sample of well-studied clusters of the LoCuSS sample \citep{Mulroy2019} at similar redshifts between $z=0.15$ and $z=0.3$. This sample provides various scaling relations between cluster observables from multi-wavelength observations of $41$ galaxy clusters and  weak lensing based mass estimates. In addition to weak lensing maps, the observations used to derive the scaling relations include X-ray, millimetre, optical, and infrared observations. The weak lensing masses inferred for the sample range from $2\times 10^{14}\,\mathrm{M_\odot}$ to $2\times 10^{15}\,\mathrm{M_\odot}$. Note that even though we use the weak lensing masses as baseline for our comparison, here they also still carry significant systematic uncertainties \citep{Ardila2021}. The mass values $M_\mathrm{500}$ for the galaxy clusters in MTNG740 are computed based on spherical overdensity measurements around the potential minimum of the galaxy cluster.

The top left panel of Figure~\ref{fig:scalingrelations} shows the richness $\lambda$. For the MTNG740 galaxy clusters we compute the richness as the number of galaxies within $R_{500}$ of the galaxy cluster with an intrinsic luminosity in the restframe $r$-band brighter than $L_\mathrm{r}<L^*_\mathrm{r}+1.75=-19.75$. Here we assume $L^*_\mathrm{r}=-21.5$ \citep{MonteroDorta2009} to mimic the luminosity cut used to compute the richness for the observed galaxy clusters. At $z=0.25$ this roughly corresponds to a stellar mass of $5\times10^9\,\mathrm{M_\odot}$ for satellite galaxies in MTNG740, which lies in the regime where the stellar mass of galaxies in MTNG740 is already significantly below the expected stellar-mass halo-mass relation, as shown in the top left panel of Figure~\ref{fig:tngmtng}. The slope of the richness-mass relation of the MTNG740 clusters agrees reasonably well with observations. The normalisation of the richness of the MTNG740 clusters at fixed cluster mass is, however, about a factor of two lower than for the observed clusters.

The interpretation of this difference is not obvious, because it could be caused either by a difference in the richness in the MTNG model or the observational measurement, or by a bias in the weak lensing mass estimate of the observed clusters. A reason for a difference in the richness estimate might be the smaller number of galaxies in MTNG740 compared to observations around and below the knee of the galaxy luminosity function, as shown in the top center panel of Figure~\ref{fig:tngmtng}. Moreover, not including a model for dust may bias the galaxy luminosities of the MTNG740 cluster galaxies. In contrast to the observational method, our richness estimate depends strongly on the exact value of the luminosity cut. If we alleviate the luminosity cut to $L_\mathrm{r}<-18$ instead, we can make the richness scaling relation match with the observed data points. On the other hand, if the weak lensing masses underestimate the true mass of the observed galaxy clusters by a factor of two, both relations will also match. However, this would also affect the other scaling relations that agree significantly better (see below) and therefore appears to be much less likely explanation. Plausibly a combination of several factors adds up to the discrepancy we see.

In the top center panel of Figure~\ref{fig:scalingrelations}, we show the Sunyaev–Zeldovich (SZ) scaling relation. We compute the integrated Compton-$y$ parameter in a sphere of radius $R_{500}$ as 
\begin{equation}
\label{eq:Y500}
Y_{500} = \frac{\sigma_\mathrm{T}}{m_\mathrm{e}c^2}\int_0^{R_{500}} \mathrm{d}V P_\mathrm{e} = \frac{(\gamma-1)\sigma_\mathrm{T}}{m_\mathrm{e}c^2} X_\mathrm{e} X \mu E_\mathrm{gas},
\end{equation}
where $\sigma_{\rm T}$ is the Thomson cross-section, $m_\mathrm{e}$ is the electron mass, $c$ is the speed of light, and $P_\mathrm{e}$ is the electron pressure, $\gamma=5/3$ is the adiabatic index, $X_\mathrm{e}=n_\mathrm{e}/n_\mathrm{H}=1.158$ is the electron-to-hydrogen number density fraction for a hydrogen mass fraction of $X=0.76$, and $\mu=0.588$ is the mean molecular weight for a fully ionised medium with primordial abundances. Here $E_\mathrm{gas}$ is the total thermal energy of the gas within $R_{500}$. For comparison with the observed clusters we show $D^2_\rmn{A} Y_{500}$, where $D_\rmn{A}$ is the angular diameter distance at $z=0.25$ for the MTNG740 cosmology. We find good agreement between MTNG740 and the observed galaxy clusters for both the shape and the normalisation of the SZ scaling relation. The MTNG740 clusters show significantly less scatter around the relation than the observed clusters. We postpone a detailed study about the origin of the scatter \citep{Battaglia2012} and the reason for the deviations from the observational scatter to future work.

The top right panel of Figure~\ref{fig:scalingrelations} shows the scaling relation for the core-excised bolometric X-ray luminosity. Detailed X-ray mocks are in principle possible \citep[see, e.g.][for TNG300 clusters]{Pop2022}, but are far beyond the scope of this paper. Here we compute the bolometric X-ray emissivity of a single cell $j$ following \citet{Kannan2016,TNGMarinacci} as
\begin{flalign}
    L_{j,\mathrm{X}} = & 6.8\times 10^{-38} \rmn{erg~K}^{1/2}\rmn{cm}^3 \frac{k_\mathrm{B}T_j^{1/2}}{h} \nonumber\\
    & \times \left[ \exp{\left(-\frac{E_\mathrm{low}}{k_\mathrm{B}T_j}\right)} - \exp{\left(-\frac{E_\mathrm{high}}{k_\mathrm{B}T_j}\right)} \right] \nonumber\\
    & \times Z^2 g_\mathrm{ff} \frac{X_\mathrm{e}}{\left(X_\mathrm{i}+X_\mathrm{e}\right)^2} \left( \frac{\rho_j}{\mu_j m_\mathrm{p}} \right)^2 V_j ,
\end{flalign}
where $T$ is the temperature in Kelvin, $h$ is Planck's constant (not to be confused with the dimensionless value of the Hubble constant used elsewhere in our work), $\rho$ the mass density and $V$ the volume of a cell (both measured in cgs units). We use $Z=\sqrt{1.15}$, $g_\mathrm{ff}=1.3$, $X_\mathrm{i}=1.079$ and $X_\mathrm{e}=1.16$ for a fully inonized gas of primordial composition \citep{TNGMarinacci}. We set the integration limits to $E_\mathrm{low}=0.7\mathrm{keV}$ and $E_\mathrm{high}=10\mathrm{keV}$ to match the observations \citep{Mulroy2019}.

Similar to \citet{Pop2022}, we exclude cells that have a temperature below $10^5\,\mathrm{K}$, star-forming cells, and cells that have a negative net cooling rate (i.e.~that are heated). Our estimated X-ray luminosity scaling relation is roughly consistent with observations. The MTNG740 clusters lie on the observed scaling relation, but with significantly smaller scatter. A significant part of the scatter is likely caused by projection effects that become visible when the simulations are better modelled via mock observables \citep{Pop2022}.

In the bottom left panel of Figure~\ref{fig:scalingrelations}, we show the scaling relation for the total $K$-band luminosity within $R_{500}$. We compute the total rest-frame $K$-band luminosity of the MTNG740 clusters as the sum of the individual $K$-band luminosity of all stars in $R_{500}$. We assume a solar luminosity in the $K$-band of $M_{{\rm K},\odot}=3.39$ \citep{Johnson1966} and do not include any dust model accounting for intrinsic dust in the galaxies and galaxy cluster. The observed data points are computed as the sum of the $K$-band luminosities of all galaxies of the clusters within a projected radius of $R_{500}$ that are consistent with being a member of the cluster and brighter than an observational detection threshold. The observed $K$-band luminosities are corrected for Milky Way dust. Both scaling relations agree well with each other in normalisation and slope. The scatter of the observed clusters is slightly larger than for the MTNG740 clusters.

The good agreement of the total $K$-band luminosity indicates that the total stellar mass in $R_{500}$ matches observations well. Since the richness, however, seems to be smaller than observed, this might indicate that galaxies in $R_{500}$ are stripped too quickly in the simulation, so that they contribute the correct stellar mass into $R_{500}$, but at any given time the population of galaxies within $R_{500}$ is fainter than in observations. Note that satellite stripping is known to depend on numerical resolution with lower resolution simulations showing faster stripping of galaxies when falling into more massive halos \citep[see, e.g.][]{Green2021}.

The bottom center panel of Figure~\ref{fig:scalingrelations} shows the scaling relation for the $K$-band luminosity of the central brightest cluster galaxy. For MTNG740 we use the integrated K-band luminosity of all stars in a fixed physical radius of $30\,\mathrm{kpc}$ as a proxy for the luminosity of the central brightest cluster galaxy. The brightest cluster galaxies in MTNG740 seem to be brighter in the $K$-band than the observed brightest cluster galaxies by about a factor of two. Note that the scatter and the slope of the scaling relations are roughly consistent with observations. The difference is interesting because it might point to shortcomings of the AGN model and its coupling to the central BCG or excessive mergers of the central galaxy with satellite galaxies. In particular, an inaccurate centering of the black hole has been argued to lead to too massive BCGs \citep{RagoneFigueroa2018}. Owing to our crude definition of the brightness of the simulated brightest cluster galaxies we refrain from a detailed investigation of the differences here. We leave this to future work that should be based on a faithful forward modelling of MTNG740, and involve mock images of the simulated clusters including a model for intrinsic dust. 

Finally, we show the gas fractions relative to the cosmic baryon fraction in the lower right panel of Figure~\ref{fig:scalingrelations}. For the simulations we compute the gas fraction as the total gas mass in $R_{500}$ divided by total mass in $R_{500}$ (i.e. $M_{500}$). The panel is essentially a zoom-in of the bottom center panel of Figure~\ref{fig:tngmtng} with a different set of data points. The gas fractions of the MTNG740 clusters lie right in the middle of the distribution of the observed data points, but the relation is much tighter for the MTNG740 clusters than for the observed clusters.

For most of the scaling relations we find that the scatter in the observed relations is significantly larger than for the MTNG740 clusters. One potential avenue to explain the differences is observational uncertainties. These contribute to the mass estimate as well as to the observed quantity of the scaling relation, but the error bars usually only include statistical contributions and at best the well-known systematic errors. Moreover, projection effects can play a significant role for the scatter of various cluster observables \citep{Pop2022,Debackere2022}. Nevertheless, it is also possible that at least some part of this scatter at fixed mass is real and reflects physics not adequately captured in our simulation. In this case, qualitative changes to our physics model will likely be required to significantly increase the object-to-object scatter at fixed mass, for example for the gas fraction or Compton-$y$ parameter. We emphasize that both better observations and more realistic comparisons between simulations and observations are required to better understand all the sources of the scatter.

\begin{figure}
\includegraphics[width=0.97\linewidth]{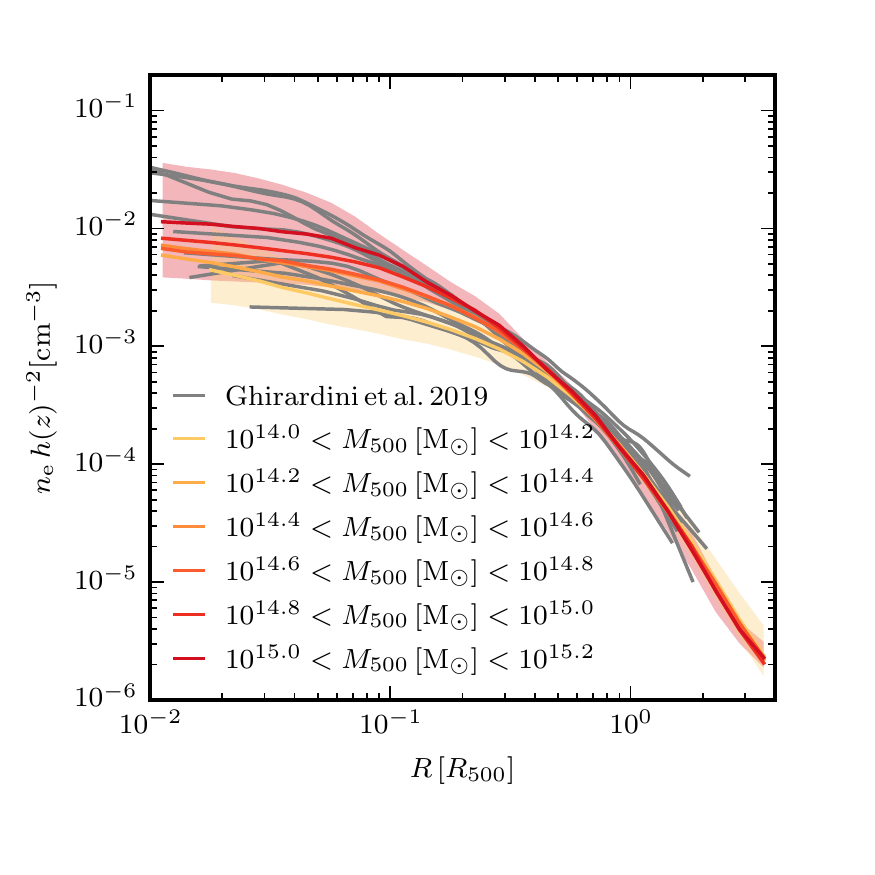}
\caption{Median profiles of the electron density of galaxy clusters with $M_\mathrm{500c} > 10^{14}\,\mathrm{M_\odot}$ at $z=0$ in various narrow mass bins, as labelled. The bands show $10\%$ and $90\%$ percentiles for the lowest and highest mass bins. We use $h(z)=H(z)/H_0$ for the MTNG740 cosmology. For comparison, we show the profiles of observed galaxy clusters from the X-COP project in grey \citep{Ghirardini2019}. We see good agreement between the MTNG740 clusters and observations. The MTNG740 clusters tend to have a lower central electron density compared to the observed cluster profiles. More massive clusters have a higher central electron density at fixed relative radius in MTNG740.}
\label{fig:profiles_ne}
\end{figure}

\section{Galaxy cluster profiles}
\label{sec:clusterprofiles}

After discussing integrated quantities of galaxy clusters, we now go one step further and focus on their internal radial profiles. We select all $2359$ clusters at $z=0$ with $M_\mathrm{500} > 10^{14}\,\mathrm{M_\odot}$ and compare them to observed profiles of nearby galaxy clusters from the X-COP project \citep{Eckert2019, Ettori2019, Ghirardini2019}. X-COP provides thermodynamic profiles as well as metallicity and stellar mass profiles \citep{Ghizzardi2021} for $12$ nearby galaxy clusters at redshifts $0.04 < z < 0.1$, with estimated masses of $4\times 10^{14}\,\mathrm{M_\odot} \lesssim M_\mathrm{500} \lesssim 10^{15}\,\mathrm{M_\odot}$. The results are based on deep XMM-Newton X-ray as well as millimeter observations to reconstruct high precision profiles of the cluster gas. The $12$ galaxy clusters of the X-COP survey are selected from the Planck all-sky SZ map and constitute the most significant detections in the corresponding redshift range.

\begin{figure}
\includegraphics[width=0.97\linewidth]{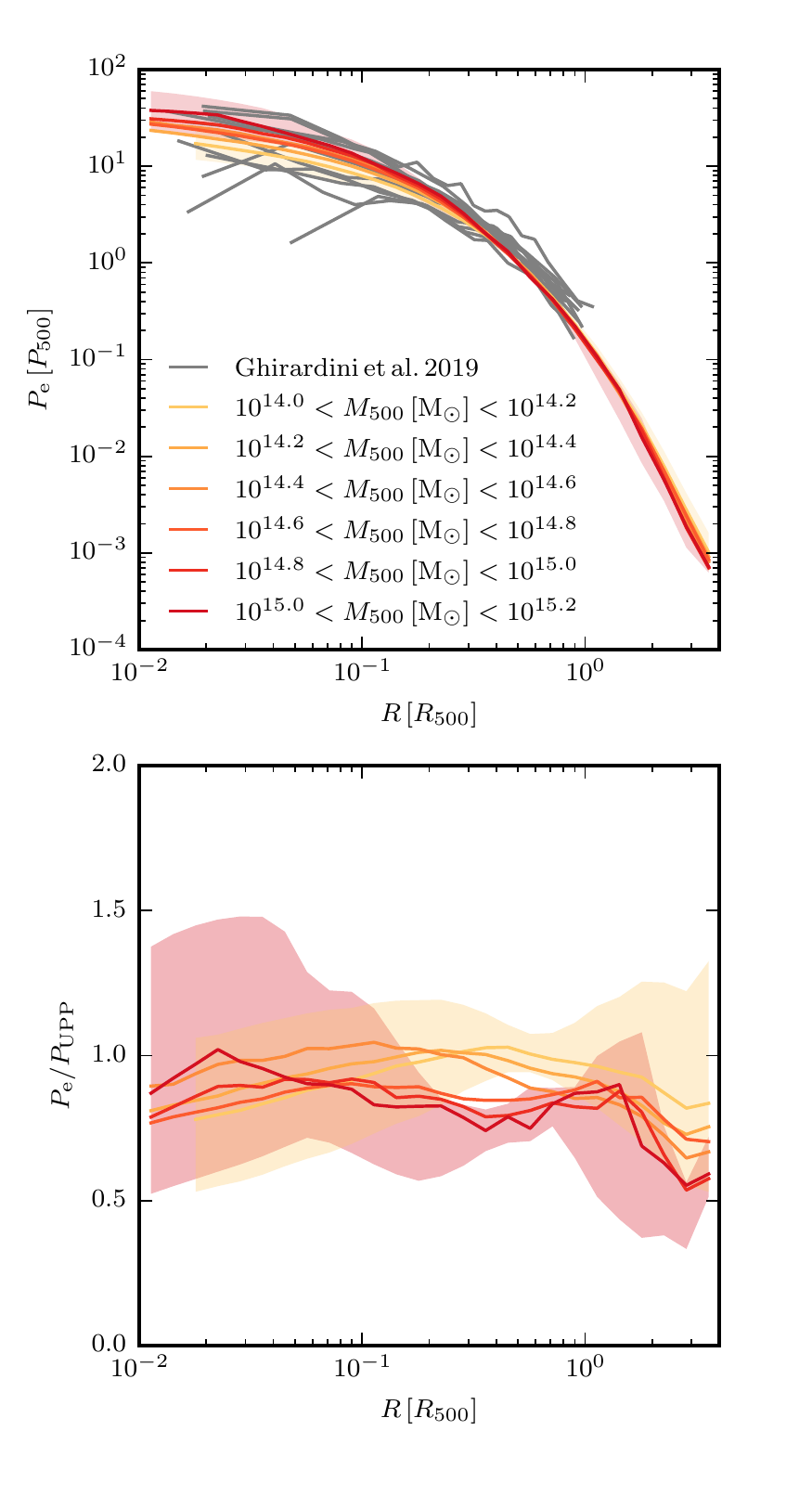}
\caption{Median pressure profiles of galaxy clusters with $M_\mathrm{500c} > 10^{14}\,\mathrm{M_\odot}$ at $z=0$ in bins of similar mass (top panel). The bands show $25\%$ and $75\%$ percentiles for the lowest and highest mass bins. We show the profiles of observed galaxy clusters from the X-COP project in grey for comparison \citep{Ghirardini2019}. The bottom panel gives the MTNG740 cluster profiles divided by the universal pressure profile for galaxy clusters \citep{Arnaud2010}. The shape of the profiles of the MTNG740 clusters agrees well with the shape of the universal pressure profile up to $R_{500}$. For massive clusters in MTNG740, the normalisation of the profile is $\sim20\%$ lower at $R_{500}$ than predicted by the universal pressure profile.}
\label{fig:profiles_pressure}
\end{figure}

We first look at different thermodynamical profiles of the gas in clusters. Even though these profiles are strongly correlated, they allow us different insights into the thermodynamical state of the gas in the MTNG740 galaxy clusters.

We start with the electron density profile in Figure~\ref{fig:profiles_ne}. We compute the volume-averaged electron density profile in spherical shells and group the MTNG740 clusters into six logarithmic mass bins in the range $14 < \log_{10}( M_\mathrm{500} / \rmn{M}_\odot ) < 15.2$. We scale the profiles with $R_{500}$ of each cluster, then combine them. We then show the median electron density profile in each mass bin with $80\%$ percentile bands, and we compare to the observed cluster profiles from the X-COP project \citep{Ghirardini2019}. All cluster profiles compare the 3D profiles of the MTNG740 clusters with the 3D profiles inferred from observed 2D profiles. We find a reassuring agreement between MTNG740 and the X-COP profiles. The typical scatter of the MTNG740 cluster profiles around the median profile is a factor of a few in the center, and almost zero at radii equal or larger than $R_{500}$. The central electron densities of the median profiles of the MTNG740 clusters tend to be slightly smaller compared to the observed clusters. However, the most extreme clusters of MTNG740 more than cover the range of the observed profiles. The central electron density of MTNG740 clusters increases systematically with cluster mass at fixed scaled radius, but reaches the same value independently of cluster mass in the outer parts, for $R\gtrsim 0.4 R_{500}$, with very small scatter.

\begin{figure}
\includegraphics[width=0.97\linewidth]{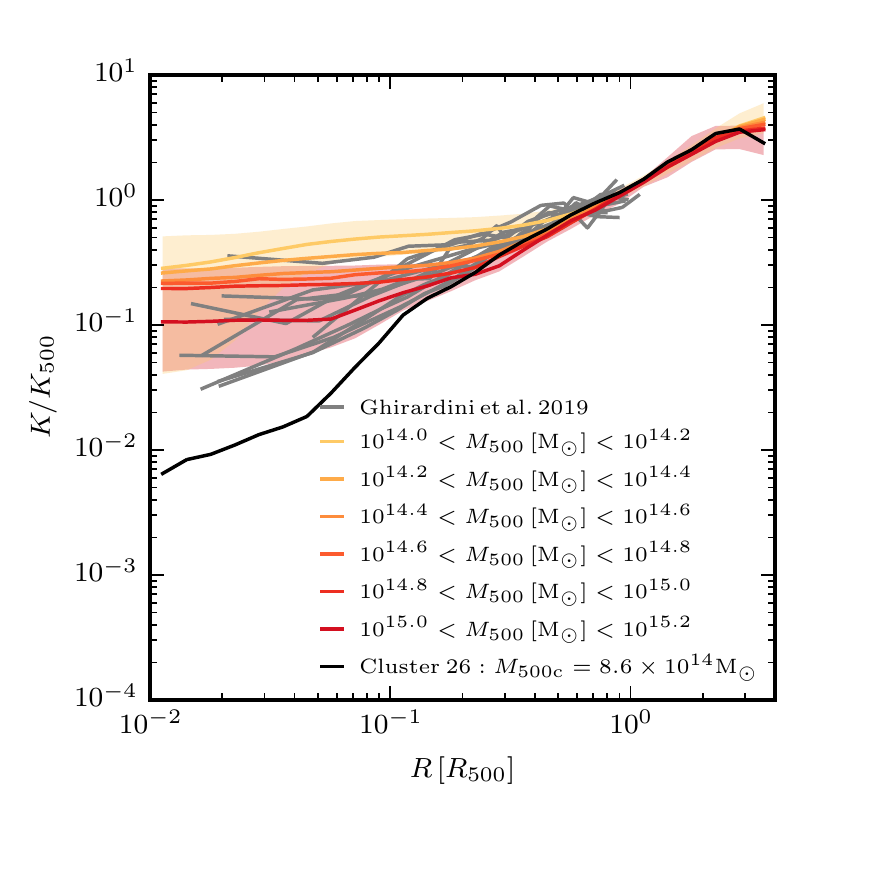}
\caption{Median entropy profiles of galaxy clusters with $M_\mathrm{500} > 10^{14}\,\mathrm{M_\odot}$ at $z=0$ in bins of similar total mass. The bands show $10\%$ and $90\%$ percentiles for the lowest and highest mass bins. The black line shows the profile of one extreme individual cool core cluster with $K<30\,\rmn{keV~cm}^2$ for $R/R_{500}<0.05$.} We compare to the entropy profiles of observed galaxy clusters from the X-COP project in grey \citep{Ghirardini2019}. We find satisfactory agreement between the MTNG740 clusters and the observed profiles. More massive clusters in MTNG740 have, however, systematically lower entropy at fixed relative radius.
\label{fig:profiles_entropy}
\end{figure}

We consider pressure profiles in the top panel of Figure~\ref{fig:profiles_pressure}, relative to $P_{500}$ defined as 
\begin{equation}
    P_{500} = 1.65\times10^{-3}h(z)^{8/3}\left[ \frac{M_{500}}{3\times10^{14}\,h^{-1}_{70}\mathrm{M_\odot}} \right]^{2/3} h^{2}_{70}~\mathrm{keV\,cm^{-3}},
\end{equation}
where $h_{70}=H_0/(70~\rmn{km~s}^{-1}\rmn{Mpc}^{-1})$ and $h(z)=H(z)/H_0$, i.e.~$h(z=0)=1$ \citep{Nagai2007}. We compute the median pressure profile in the same way as the profile of the electron density, and again compare to the clusters of the X-COP survey \citep{Ghirardini2019}. We find good agreement between the profiles of the MTNG740 clusters and the observed clusters. There is no systematic trend with cluster mass after expressing the pressure in terms of $P_{500}$. The scatter in the profiles of the MTNG740 clusters around $R_{500}$ is slightly smaller than in the observations, but the latter is likely dominated by observational systematics.

In the bottom panel of Figure~\ref{fig:profiles_pressure} we divide the pressure profile by the universal pressure profile \citep{Arnaud2010} that is often assumed when converting cluster observables to physical properties of the cluster. The electron pressure profiles of the MTNG740 clusters are well described by the universal pressure profile out to $R_{500}$, with deviations smaller than $20\%$. However, at larger radii the universal pressure profile systematically underestimates the pressure in the MTNG740 clusters.

We consider the entropy profile in Figure~\ref{fig:profiles_entropy}. For the MTNG740 clusters, we define the entropy as $K=k_\rmn{B} T / n_\mathrm{e}^{2/3}$ \citep{Henden2018} and express it in units of the characteristic entropy $K_{500} = k_\rmn{B} T_{500} / n_\mathrm{e,500}^{2/3}$, using
\begin{equation}
    T_{500} = \frac{\mu m_\rmn{p} G M_{500}}{2 k_\rmn{B} R_{500}}
\end{equation}
and
\begin{equation}
    n_\mathrm{e,500} = \frac{500 f_\rmn{b} \rho_\mathrm{c}}{\mu_\rmn{e} m_\rmn{p}},
\end{equation}
where $k_\rmn{B}$ is the Boltzmann constant, $m_\rmn{p}$ is the mass of the proton, $f_\rmn{b}=\Omega_\rmn{b} / \Omega_\rmn{M}$ is the baryon fraction of the Universe, and $\rho_\mathrm{c}$ is the critical density of the Universe at $z=0$. We again compare to the profiles from the X-COP survey. We find that the median central entropy is significantly lower for more massive clusters in MTNG740. Overall, the entropy profiles of the MTNG740 clusters agree reasonably well with the observed profiles, but the median central entropy of the simulated clusters is systematically higher than the observed central entropy profiles. This can partly be understood as reflecting the differences in the central electron density profile that is lower for the MTNG740 clusters compared to the observed clusters. The kinetic AGN feedback in TNG quickly dissipates and increases the central entropy on the sound crossing time scale. As a consequence, the cluster population can mostly be described by non-cool core clusters (with central entropies $>30\,\rmn{keV~cm}^2$), even though a small number of cool core clusters exists in MTNG740. Nevertheless, MTNG740 misses out on a large fraction of cool-core clusters \citep[as shown for the TNG300 sample by][]{Barnes2018}. Fitting an entropy profile of the form
\begin{equation}
    K \left( r \right) = K_0 + K_{100} \left( \frac{r}{100\,\mathrm{kpc}} \right)^a
\end{equation}
in the range $10^{-2}R_{500}-R_{500}$ \citep{Cavagnolo2009,Barnes2018} we find that $196$ or $8\%$ of all clusters at $z=0$ with $M_{500}>10^{14}M_\odot$ have $K_0<30\rmn{keV~cm}^2$ and can be classified as cool core clusters, consistent with TNG300 \citep{Barnes2018}. None of the $9$ clusters with $M_{500}>10^{15}M_\odot$ fulfill this criterion. The lack of cool core clusters is similar to purely thermal AGN feedback models \citep{Altamura2022}, in contrast to light AGN jet feedback that is able to maintain the cool core while self-regulating the cooling ICM \citep{Ehlert2022,Weinberger2022}.

\begin{figure}
\includegraphics[width=0.97\linewidth]{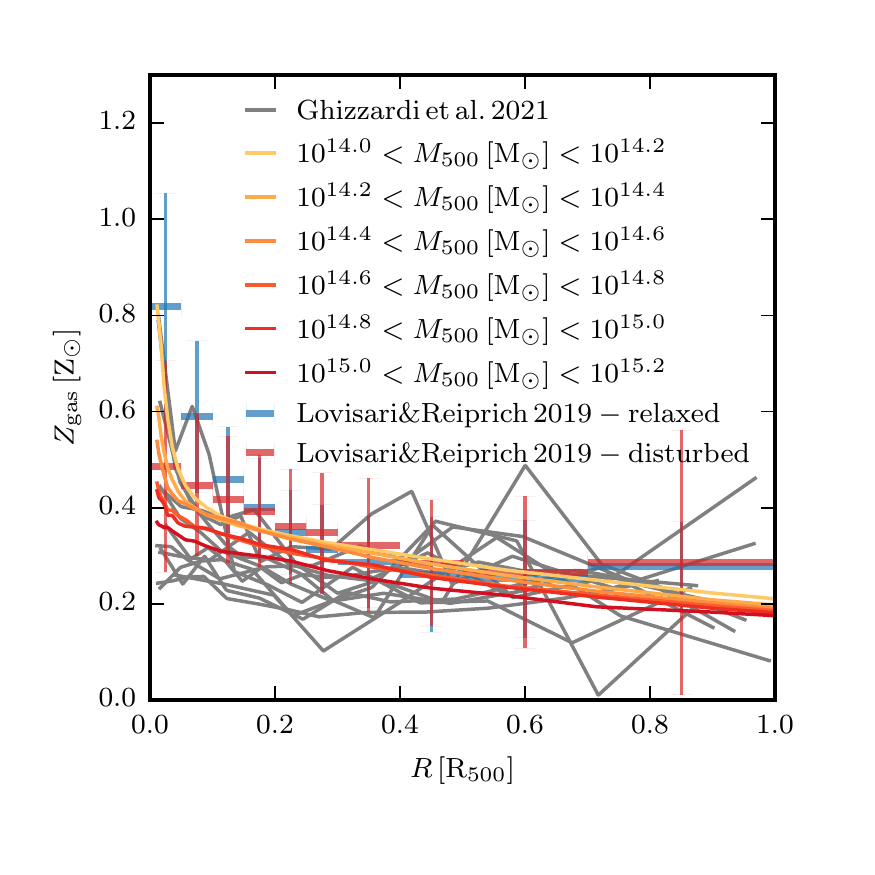}
\caption{Median metallicity profiles of galaxy clusters with $M_\mathrm{500} > 10^{14}\,\mathrm{M_\odot}$ at $z=0$ in bins of similar mass for a solar metallicity of $Z_\mathrm{\odot}=0.0127$. For comparison, we show observed profiles of two sets of nearby galaxy clusters \citep{Lovisari2019,Ghizzardi2021}. The metallicity profiles of the MTNG740 galaxy clusters are in broad agreement with observed metallicity profiles of galaxy clusters.}
\label{fig:metallicity}
\end{figure}

In Figure~\ref{fig:metallicity} we look at the metallicity profiles of the intracluster gas. For the MTNG740 clusters we first compute the mass-weighted average metallicity in spherical shells for each cluster, then combine the profiles to obtain median profiles in six mass bins. We compare to the observed metallicity profiles by \citet{Lovisari2019} and \citet{Ghizzardi2021}. We note that the observations typically measure iron abundance as a proxy for the metallicity of the gas in galaxy clusters.  However, we only track the total metal abundance in MTNG740 (forced by memory constraints), and thus use it for the comparison. The median metallicity profiles of MTNG740 clusters for different mass bins are very similar. They show only a small but still systematic trend with the mass of the clusters. Specifically, more massive galaxy clusters have slightly lower gas metallicity at fixed relative radius. This is consistent with basic expectations because the fraction of baryons that is converted to stars and eventually produces metals decreases with halo mass (see also the top right-hand panel of Figure~\ref{fig:tngmtng}). We find overall good agreement with the observed metallicity profiles. The inner slope of the observed clusters, most notably for the sample of relaxed clusters, seems to be slightly steeper than for the MTNG740 clusters, though a similar trend is not obvious in the X-COP sample. We leave a more detailed analysis that splits the MTNG740 galaxy clusters into relaxed and disturbed clusters in a similar way as done for observed clusters to future work.

\begin{figure}
\includegraphics[width=0.97\linewidth]{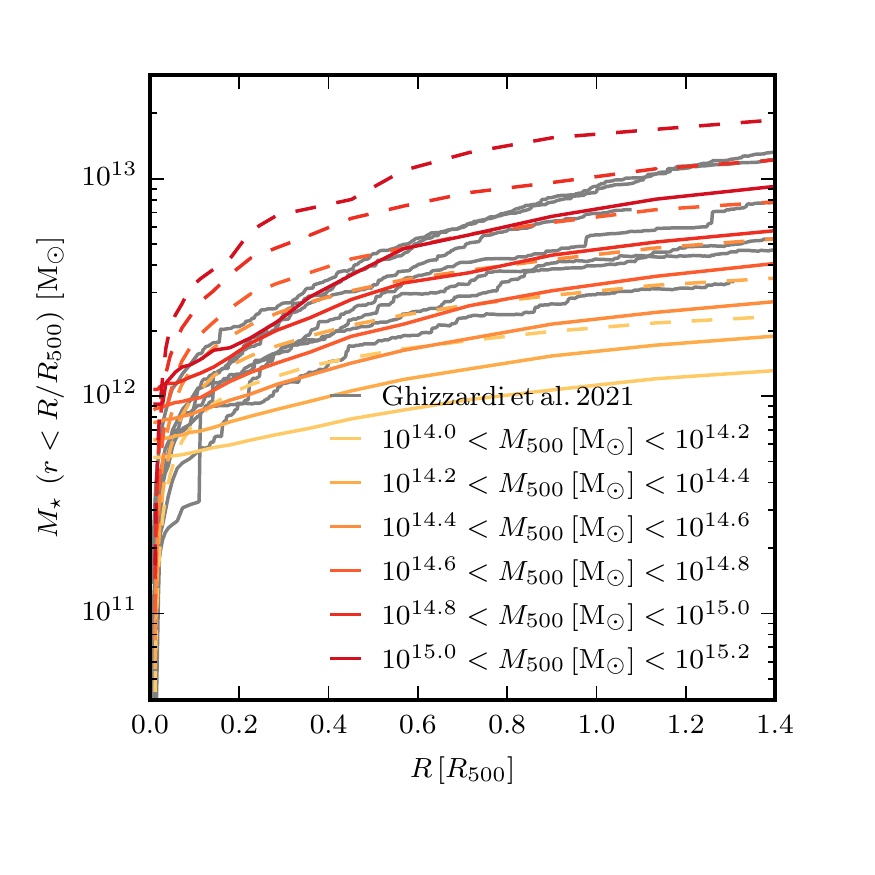}
\caption{Median cumulative stellar mass profiles of galaxy clusters with $M_\mathrm{500c} > 10^{14}\,\mathrm{M_\odot}$ at $z=0$ in bins of similar mass. Solid lines show the stellar mass within a certain radius in the BCG and satellite galaxies, dashed lines show the total stellar mass within a certain radius including intracluster light. The bands show $10\%$ and $90\%$ percentiles. For comparison we show observed profiles of nearby galaxy clusters from the X-COP project computed from all galaxies in the clusters \citep{Ghizzardi2021}. The BCG and satellite galaxy based stellar mass profiles are in good agreement with the observed profiles generated the same way. The total stellar mass profiles of the MTNG740 clusters are significantly higher owing the a significant contribution of diffuse intracluster light.}
\label{fig:stellarmass}
\end{figure}

Finally, we show cumulative stellar mass profiles in Figure~\ref{fig:stellarmass} and again compare to the observed profiles of the clusters of the X-COP project \citep{Ghizzardi2021}. The stellar mass profiles of the MTNG740 clusters computed from their BCG and satellite galaxies look  consistent with the stellar mass profiles of the observed clusters that are generated in a similar way. The observed stellar mass profiles are slightly steeper at large radii.

The total stellar mass profiles of the MTNG740 clusters, however, are significantly higher showing that there is a significant amount of stellar mass in a diffuse intracluster light component. This component is missed when only galaxies in the cluster are counted. Interestingly the total K-band luminosity of the MTNG740 clusters seems to agree well with the scaling relations shown in Section~\ref{sec:scalingrelations} even though the total K-band luminosity of the observed clusters was also calculated from cluster galaxies. We leave this apparent discrepancy for future work.

\begin{figure*}
\includegraphics[width=0.97\linewidth]{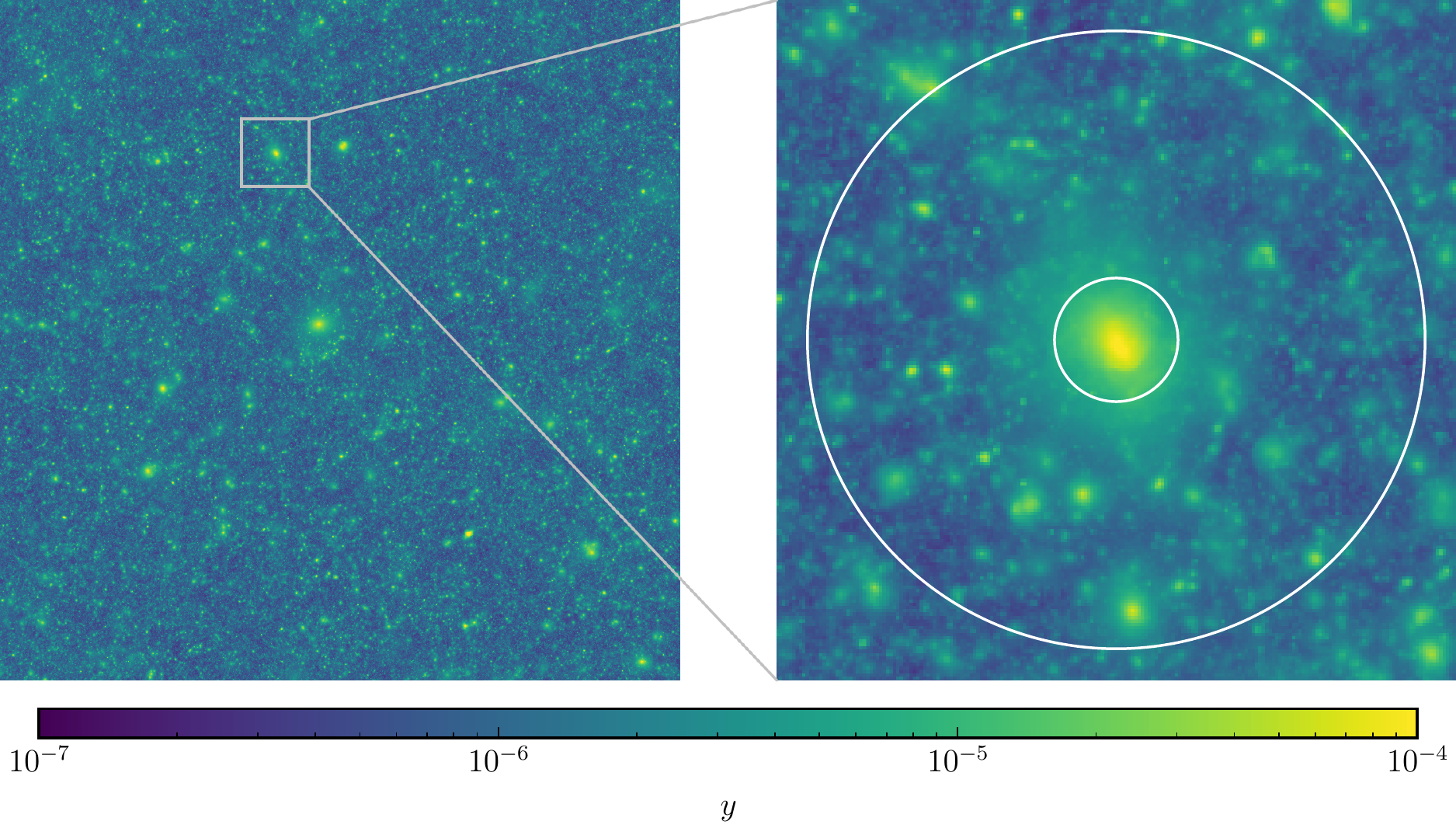}
\caption{Map of the Compton-$y$ parameter for a deep $10^\circ\times10^\circ$  lightcone that extends to $z=5$. The right panel shows a zoom-in on the largest galaxy cluster with mass $M_{500}=7\times10^{14}\,\mathrm{M_\odot}$ at $z=0.25$ contained on this lightcone. The white circles show $R_{500}$ and $5\,R_{500}$ of this cluster.}
\label{fig:lightconeMap}
\end{figure*}

We find that the median cumulative stellar mass profiles are almost self-similar for different mass bins, with a normalisation that increases with cluster mass. There is reasonable agreement of the outer shape of the stellar mass profiles between MTNG740 and observations. However, the inner stellar mass profiles are significantly steeper for the MTNG740 clusters, and their stellar mass is more centrally concentrated than for the observed profiles. The total stellar mass in the cluster is larger in MTNG740 clusters, but the stellar mass in galaxies in MTNG740 clusters is lower than observed. This difference indicates that galaxies are stripped more quickly in MTNG740 than in the real Universe, which can increase the intracluster light contribution in MTNG740 at the expense of the individual galaxies. However, \citet{Ahad2021} argue that, if anything, galaxies are stripped too little in the Hydrangea simulations, which have a roughly similar mass resolution as MTNG740. We compare the properties of the cluster galaxies in MTNG740 with observations in more detail in Figure~\ref{fig:clustergalaxies} and discuss these results further in Section~\ref{sec:observations}. Nevertheless, as shown in previous simulations, the intracluster light can contribute a substantial fraction of the stellar mass outside the central brightest cluster galaxy \citep{Puchwein2010}. A similar result has later been found in observations with sufficiently deep surface brightness sensitivity \citep{Presotto2014,deOliveira2022}. Future work should mimick the stellar light of the simulated MTNG740 clusters and apply the same observational cuts and limitations to facilitate a proper comparison.

\section{The SZ-signal on the lightcone}
\label{sec:lightcones}

While spherical profiles and integrated quantities on spherical apertures greatly assist physical understanding, these measures cannot be observed directly. Instead, we can only access projections of an object on the sky. More strictly speaking, we only see our past backwards lightcone. To understand the relevance of this difference for SZ observations, we first consider the Compton-$y$ parameter for a full, deep lightcone output of our simulation in Figure~\ref{fig:lightconeMap}. This lightcone extends to redshift $z=5$ and covers a square area of $10^\circ\times10^\circ$ on the sky. 

We compute the Compton-$y$ of a pixel on the lightcone in a similar way as $Y_{500}$ in Equation~\eqref{eq:Y500} from the total thermal energy of the electrons in the volume covered by pixel on the sky as
\begin{equation}
    y\left(\theta \right) = \frac{k_\mathrm{B}\sigma_\mathrm{T}}{m_\mathrm{e}c^2}\int_0^{D} n_\mathrm{e} T_\mathrm{e} \mathrm{d}l = \frac{(\gamma-1)\sigma_\mathrm{T}}{m_\mathrm{e}c^2} \widetilde{x}_\mathrm{e} X \mu \frac{1}{\Omega} \sum_i \frac{E_i}{D^2_{\mathrm{A},i}},
\end{equation}
where $E_i$ is the thermal energy of a cell and $D_{\mathrm{A},i}$ the angular diameter distance at the redshift when the cell crosses the lightcone, $\Omega$ is the solid angle of the pixel on the sky and $\Sigma_i$ includes all lightcone cells whose centers lie on the pixel.

Massive galaxy clusters clearly stick out with a high central Compton-$y$ up to $10^{-4}$. We find a mean background value on the full lightcone map of  $\bar{y}=1.68\times10^{-6}$ (mean) and $\bar{y}=1.12\times10^{-6}$ (median). This background level is still consistent with the upper bound from Planck after removing galaxy clusters with $\bar{y}<2.2\times10^{-6}$ \citep{Khatri2015}. Note, however, that the latter bound was computed for a slightly different cosmology. Our background level is also consistent with previous estimates from \citet{Springel2001b}, the Magneticum simulation \citep[$\bar{y}<1.18\times 10^{-6}$,][]{Dolag2016} and with estimates from semi-analytical models on top of dark matter only simulations \citep{Osato2018}.

In the right panel of Figure~\ref{fig:lightconeMap}, we zoom in on the most massive galaxy cluster at $z=0.25$ in the deep lightcone map. This galaxy cluster has a mass of $M_{500}=7\times10^{14}\,\mathrm{M_\odot}$ at $z=0.25$, and the white circles in Figure~\ref{fig:lightconeMap} denote its $R_{500}$ and $5\,R_{500}$ radii. These are typical aperture radii used to estimate the integrated Compton-$y$ parameter of galaxy clusters. Interestingly, we see a great deal of substructure projected onto the cluster, in particular between $R_{500}$ and $\,5R_{500}$, though most of this substructure is in the fore- and background and not in the immediate vicinity of the cluster, as we will show in the following.

\begin{figure*}
\includegraphics[width=0.97\linewidth]{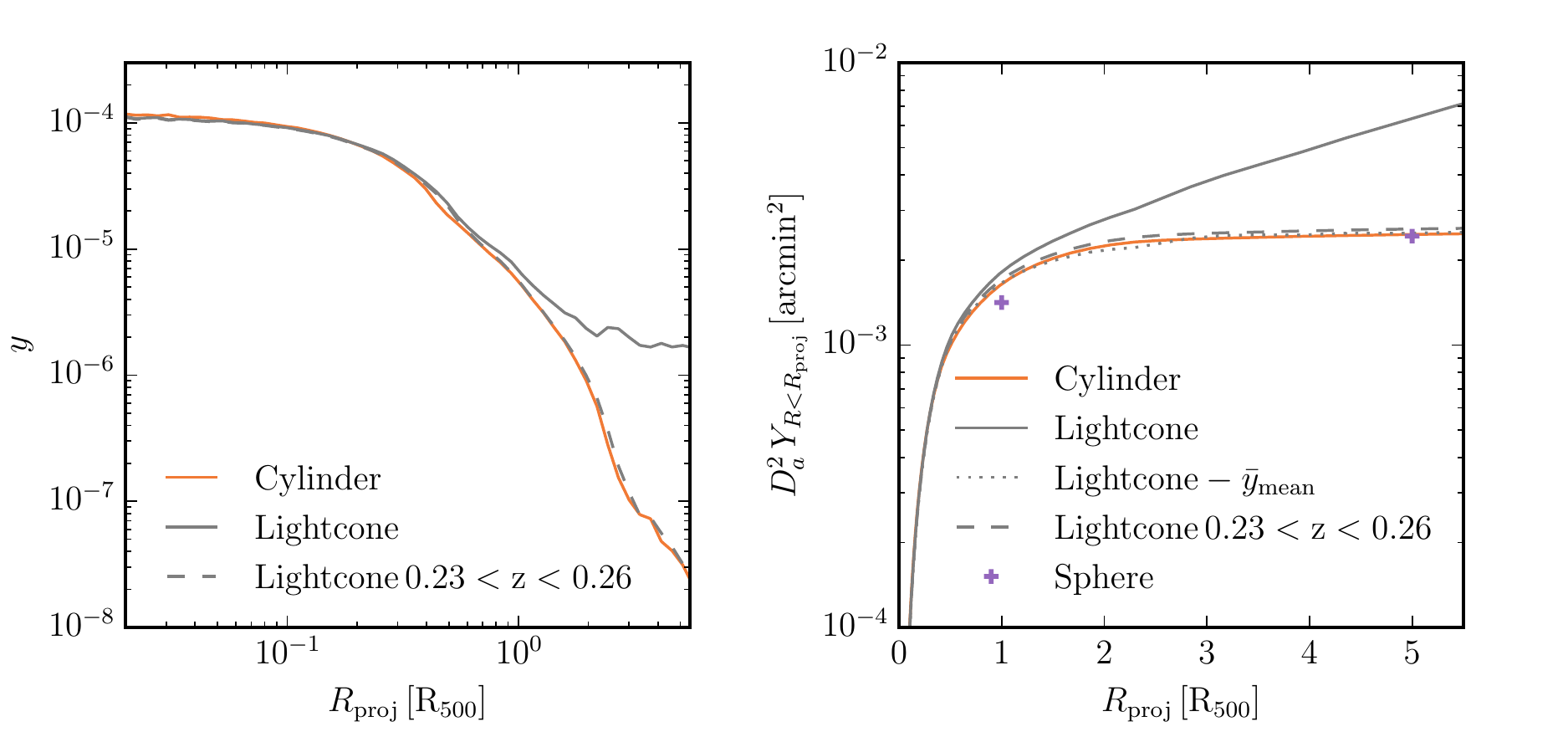}
\caption{Compton-$y$ profile of the cluster shown on the right of Figure~\ref{fig:lightconeMap} at $z=0.25$. The left panel of the present figure shows the radially averaged profile computed from the full lightcone (solid grey line) as well as the profile computed only from the cells contributing to the lightcone between $z=0.23$ and $z=0.26$ (dashed grey line). The orange line gives the same profile but computed from a cylinder with a depth of $100\,\mathrm{Mpc}$ from the snapshot data at $z=0.25$. The right panel plots the integrated Compton-$y$ parameter computed from the lightcone and cylinder projections shown in the left panel, and compares them to the integrated Compton-$y$ parameter computed for a spherical aperture at $R_{500}$ and $5\,R_{500}$. Additionally we show the integrated Compton-$y$ parameter from the full lightcone after subtracting the mean Compton-$y$ from the full lightcone shown in Figure~\ref{fig:lightconeMap}. Background subtraction is crucial to reconstruct the integrated Compton-$y$ of the cluster.}
\label{fig:lightconeCluster}
\end{figure*}

To examine this point further, we consider the differences between measuring the Compton-$y$ on the deep lightcone and from a local aperture that considers the cluster and its large-scale environment extending over 100~Mpc. We compute the projected Compton-$y$ profile of the cluster on the lightcone as shown in the left-hand panel of Figure~\ref{fig:lightconeCluster}. This is compared to the projected Compton-$y$ profile of the same cluster when seen in a cylindrical projection of the snapshot time-slice at $z=0.25$, using a total depth of $\sim100\,\mathrm{Mpc}$ around the cluster centre and the same viewing direction as for the lightcone observer. Both resulting radial profiles are contrasted in the left panel of Figure~\ref{fig:lightconeCluster}. Additionally, we also include separately a projected profile of the lightcone, but with contributions only from gas between $z=0.23$ and $z=0.26$; i.e.~to local contributions around the redshift of the cluster with a depth similar to the cylindrical projection.

We find that the profiles computed from the cylindrical aperture or from the local lightcone contributions are essentially identical. In contrast, the radial profile of the full lightcone starts to significantly deviate at a projected radius of $\sim R_{500}$, when the Compton-$y$ approaches the mean background level of the lightcone.

To quantify this difference better we show the integrated Compton-$y$ parameter for the galaxy cluster out to a given radius in the right panel of Figure~\ref{fig:lightconeCluster}. We see that $Y_{500,\mathrm{Cylinder}}$ is $10\%$ smaller than $Y_{500,\mathrm{Lightcone}}$, while $Y_{500,\mathrm{Sphere}}$ that we used in Figure~\ref{fig:scalingrelations} is $20\%$ smaller than $Y_{500,\mathrm{Lightcone}}$. 

For a radius of $5\,R_{500}$ as used by Planck \citep{PlanckSZCatalogue2016} the situation is different. $Y_{5R500}$ is essentially identical for the spherical and cylindrical apertures. Thus the local background that is included in the cylindrical aperture but not the spherical aperture does not significantly contribute to the total signal. However, $Y_{5R500}$ measured on the lightcone is $2.5$ times larger than measured for either local aperture, as the distant background that can be seen in Figure~\ref{fig:lightconeMap} contributes very significantly in this case. Hence, a careful background removal as done via matched or scale-adaptive filtering \citep{Tegmark1998,Sanz2001,Herranz2002a,Herranz2002b,Schaefer2006,Kay2012} is crucial. Future studies can use MTNG740 to test the full procedure used by SZ surveys and to identify potential systematics still present in these procedures.

\section{Galaxy cluster observables}
\label{sec:observations}

\begin{figure}
\includegraphics[width=0.97\linewidth]{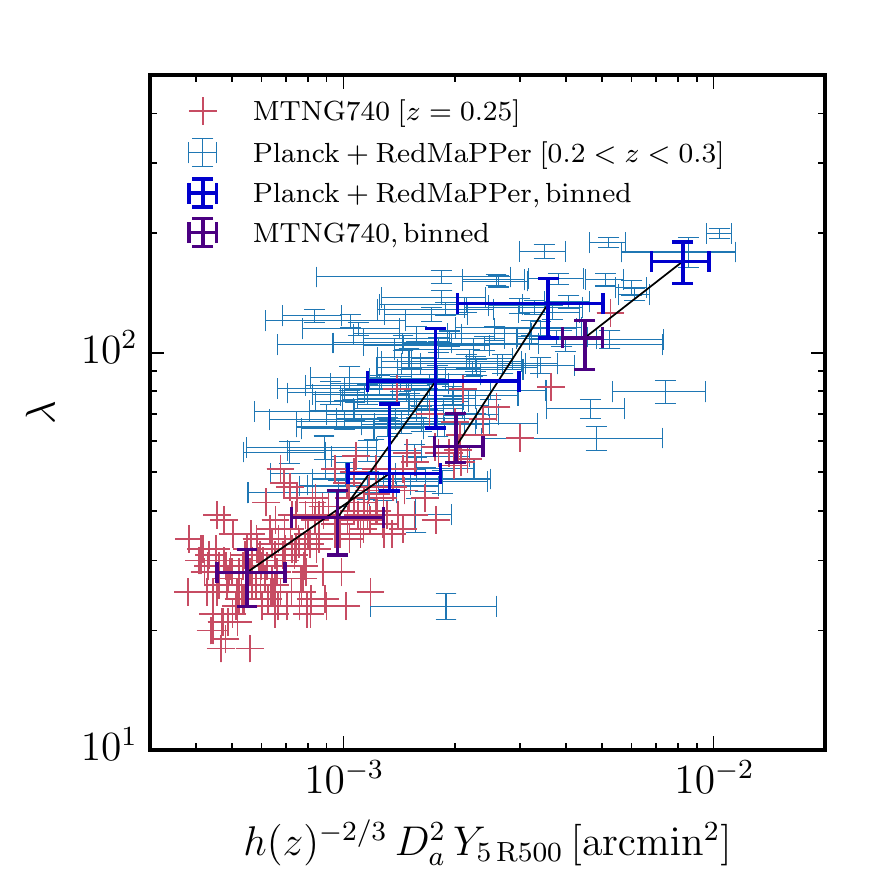}
\caption{Richness versus Compton-$y$. Light blue crosses show galaxy clusters in MTNG740 in the $z=0.25$ snapshot with $M_\mathrm{500} > 10^{14.4}\,\mathrm{M_\odot}$. Orange data points give all galaxy clusters between $z=0.2$ and $z=0.3$ that are both in the Planck SZ catalogue \citep{PlanckSZCatalogue2016,PlanckSZ2017} and the RedMaPPer SDSS-8 catalogue \citep{Rykoff2014}. We also show median values and error bars for $16\%$ and $84\%$ percentiles in four logarithmic mass bins for MTNG740 clusters (blue) and observed clusters (brown). Black lines connect points of clusters of the same mass. The individual MTNG740 clusters lie at the lower end within the parameter space of observed clusters. The binned distributions seem however to be consistent, except for a systematically lower cluster mass in MTNG740 at fixed richness and Compton-$y$. The scatter between MTNG740 clusters is smaller than the scatter between observed clusters.}
\label{fig:richness_sz}
\end{figure}

\begin{figure}
\includegraphics[width=0.97\linewidth]{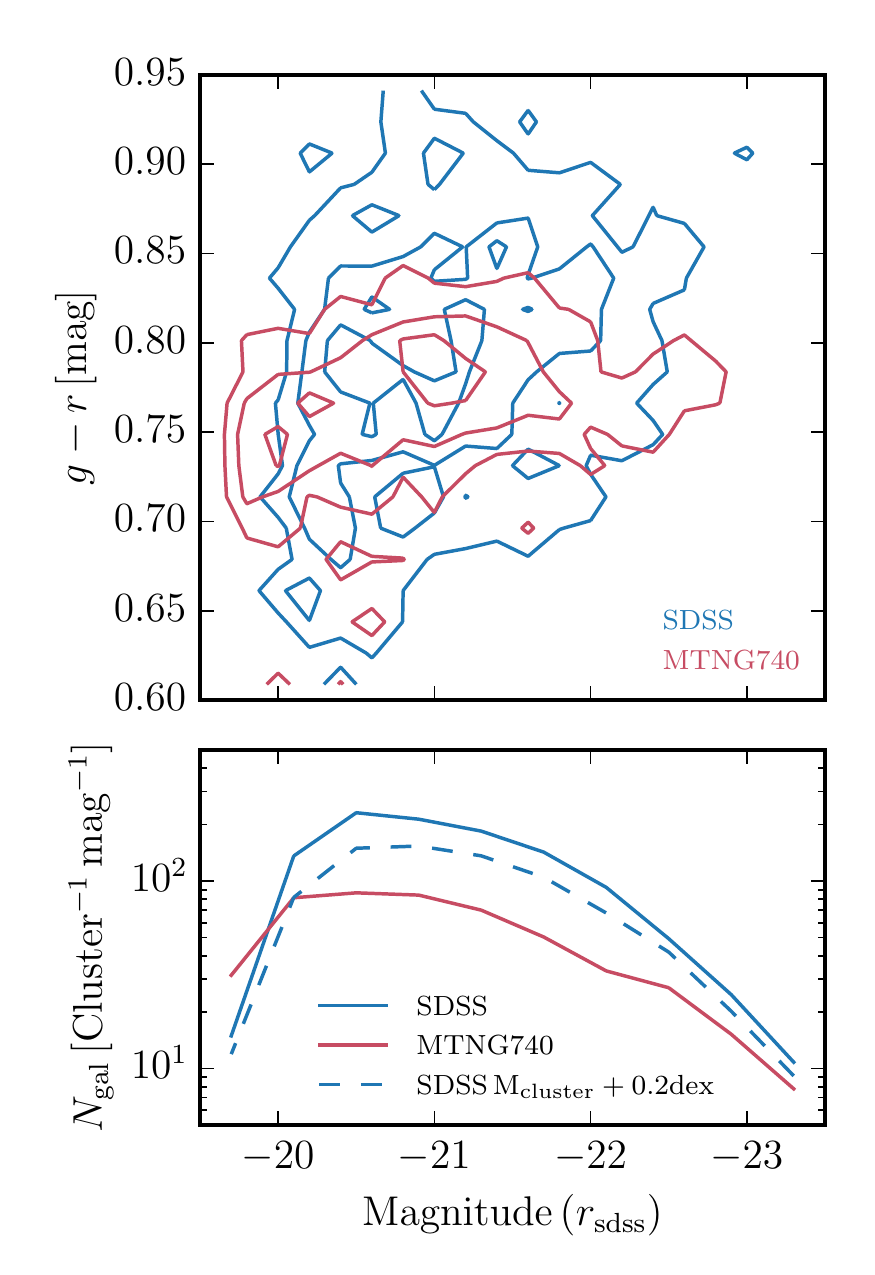}
\caption{Galaxy distributions in rest-frame $g-r$ color versus apparent rest-frame $r$-band magnitude (upper panel). We use all galaxies with an apparent $i$-band magnitude $m_i < 21$ at $z=0.25$ within $R_\mathrm{500}$ of the MTNG740 clusters shown in Figure~\ref{fig:richness_sz} that have a mass larger than $10^{14.6}\mathrm{M_\odot}$}. We compare to all galaxies that have a membership probability larger than zero for one of the clusters of the Planck+RedMaPPer sample used in Figure~\ref{fig:richness_sz}. Observed clusters are weighted with their membership probability. The color and brightness distributions of both galaxy populations are consistent, with slightly larger scatter in color in the observed galaxies. Note that no dust was included to compute the MTNG740 galaxy magnitudes. The lower panel shows the number of galaxies per cluster and magnitude in the $i$-band for both samples. The dashed line shows the observed galaxies from all Planck clusters with the same mass cut, but assuming that their masses are $0.2\,{\rm dex}$ larger. There is a clear  difference between MTNG740 and the Planck cluster galaxies that increases towards fainter galaxies. It is reduced but still there if the Planck clusters are assumed to be more massive.
\label{fig:clustergalaxies}
\end{figure}

For all comparisons of galaxy cluster properties with observations discussed so far we have assumed that we precisely know the mass of the observed galaxy clusters. However, mass is unfortunately not a direct observable for galaxy clusters, so mass estimates always require additional assumptions about the galaxy clusters that all introduce significant systematic uncertainties. 

We can avoid these assumptions by comparing to observations fully in observational space. Here, we present a first example of this approach with MTNG740 by comparing two directly observable quantities of the MTNG740 clusters with corresponding measurements made for observed galaxy clusters. For definiteness, we chose the richness and the Compton-$y$ parameter. Both quantities are almost completely independent probes of the state and properties of a galaxy cluster. We here do not attempt to constructing fully synthetic observations that are analysed in the same way as observational data, but we instead compare the theoretical quantities $Y_{5R_{500}}$; i.e.~the integrated Compton-$y$ parameter in a sphere with a radius $5\,R_{500}$, and the richness $\lambda$. Both quantities are well defined and easily determinable for the MTNG740 clusters, but they still require some modest assumptions to be made when measuring them from observations. Specifically, we compare the MTNG740 clusters to a sample of galaxy clusters that have both SZ measurements from Planck \citep{PlanckSZCatalogue2016,PlanckSZ2017}, and a richness measurement from the RedMaPPer catalogue based on the SDSS-8 galaxy catalogue \citep{Rykoff2014,Rykoff2016}.

We compare the MTNG740 clusters to the observed sample in Figure~\ref{fig:richness_sz} for all MTNG740 clusters at $z=0.25$ with minimum mass of $M_\mathrm{500} > 10^{14.4}\,\mathrm{M_\odot}$, roughly matching the detection limit of galaxy clusters in  Planck at this redshift.
We compare to the sample of clusters that are detected in Planck and that are part of the RedMaPPer SDSS8 catalogue in the redshift range $z=0.2$ and $z=0.3$. We compute the richness and integrated Compton-$y$ parameter or the MTNG740 clusters as described in Section~\ref{sec:scalingrelations}.

The MTNG740 clusters fall well within the parameter space covered by the observed galaxy clusters. The scatter at fixed richness or fixed Compton-$y$ parameter seems to be larger for the observed clusters than for the MTNG740 clusters. This difference is most easily explained by observational systematics that contribute to the scatter. The MTNG740 clusters lie at the low richness end of the distribution of observed clusters. However, this offset is much smaller than what we found for the scaling relations (see Figure~\ref{fig:scalingrelations}), which may indicate that the weak lensing mass estimates that we used for the observed clusters could be biased and be responsible for at least parts of the discrepancy.

Importantly, MTNG740 lacks galaxy clusters with the largest values of richness and Compton-$y$ found in the Planck+RedMaPPer sample. A simple explanation for this discrepancy could be that the most massive MTNG740 clusters are significantly less massive than the clusters in the observed sample due to the still limited volume of our simulation. This explanation is consistent with a rough estimate of the observational volume of SDSS-8 that covers roughly $25\%$ of the sky, which is equivalent to volume about four times larger between $z=0.2$ and $z=0.3$ than the MTNG740 volume.

To further test this hypothesis we group both sets of galaxy clusters into $4$ logarithmic mass bins from $10^{14.4}\,\mathrm{M_\odot}$ to $10^{15.2}\,\mathrm{M_\odot}$ with a width of $0.2\,{\rm dex}$ each. We find object counts of the Planck+RedMaPPer sample of $9$, $60$, $14$, and $2$, whereas the corresponding counts for the MTNG740 clusters are $191$, $72$, $19$, and $3$, for bins of increasing cluster mass. Surprisingly, we find a slightly larger number of objects in the MTNG740 cluster sample at the high mass end. However, the richness and Compton-$y$ values for MTNG740 clusters are significantly smaller than the values found for Planck+RedMaPPer clusters at a similar estimated mass. 

There are two obvious ways to explain the discrepancies. Firstly, if we assume that the SZ based mass estimate from the Planck SZ catalogue is underestimating the true mass by $\sim 0.2\,\mathrm{dex}$ we find almost perfect agreement between both samples. In this case the richness, the integrated Compton-$y$ values, as well as the number of clusters in the highest mass bins are consistent within statistical scatter between both sets of clusters after taking into account the differencs in probed volumes.

Remarkably, this factor of $\sim 0.2\,\mathrm{dex}$ to correct the mass estimates of the Planck clusters is very similar to the empirical correction to the SZ mass estimates applied in the ACT SZ catalogue \citep{Hilton2021}. Their correction is based on the observed richness of a subset of the ACT clusters, and a relation between richness and weak lensing mass estimates \citep[][]{McClintock2019}.

One problem with this interpretation is that the MTNG740 clusters do not lie on the richness scaling relation shown in Section~\ref{sec:scalingrelations}. This is not completely unexpected as MTNG740 galaxies around and below the knee of the stellar mass function are not massive enough as shown in Figure~\ref{fig:tngmtng}, even though it is not clear how much this changes the richness estimate. However, it may also indicate more complicated interpretations of the richness estimate of the MTNG740 cluster for different brightness limits of cluster galaxies or possibly an observational bias in the richness estimate of the LoCuSS sample \citep{Mulroy2019}.

As a consistency check of the richness of the MTNG740 clusters, we select all galaxies of all clusters with a mass larger than $10^{14.6}\,\mathrm{M_\odot}$ that contribute to the richness estimate of Figure~\ref{fig:richness_sz} and show their distribution in absolute restframe $r$-band magnitude versus restframe SDSS $g-r$ color space in the top panel of Figure~\ref{fig:clustergalaxies}.
For the Planck+RedMaPPer sample, we include all SDSS galaxies with a non-zero cluster membership probability and weight them with their individual probability. Moreover, we weight every galaxy with the inverse of the number of clusters in the cluster mass bin of its parent cluster. Here, we use the same four mass bins as in Figure~\ref{fig:richness_sz}. Note that the SDSS galaxies are de-reddened for Milky Way dust, but not for intrinsic dust. We apply K-corrections \citep{Chilingarian2010} to obtain rest-frame $r$-band magnitudes of the observed galaxies and use the redshift of their parent galaxy cluster to transform them to absolute magnitudes. This yields good agreement in color and brightness between the MTNG740 cluster galaxies and cluster galaxies of the Planck+RedMaPPer sample, only the scatter in color is smaller for the MTNG740 cluster galaxies. In the bottom panel of Figure~\ref{fig:clustergalaxies} we show the average number of galaxies per cluster and decade of magnitude for MTNG740 and the Planck+RedMaPPer sample. They are similar at the bright end, but there are progressively fewer galaxies in MTNG740 clusters compared to the Planck clusters the fainter the galaxies become. The lower panel of Figure~\ref{fig:clustergalaxies} also shows that this difference becomes smaller but does not vanish (dashed line) when we assume that the Planck clusters are $0.2\,\mathrm{dex}$ more massive, which changes which clusters are matched to the MTNG740 clusters. While this is mostly expected from Figure~\ref{fig:richness_sz}, it could still be explained by statistical variance.

If we alternatively assume that the SZ-based mass estimates of the Planck clusters are correct, we need to conclude that the MTNG740 clusters systematically underestimate the richness as well as the integrated Compton-$y$. The former is consistent with the richness scaling relation, but the latter would break the SZ scaling relation as shown in Figure~\ref{fig:scalingrelations}. Because Planck measures $Y_{5R_{500}}$ rather than $Y_{500}$ as shown in the SZ scaling relation, this would require significantly stronger AGN-driven outflows that heat the gas around massive galaxy clusters on scales well beyond $R_{500}$ without changing the properties of the gas within $R_{500}$. Moreover, the Planck+RedMaPPer sample would need to lack a significant number of the most massive clusters expected to be found in the survey volume.

We conclude that comparisons between simulations and observations in observational space are a crucial and promising test of simulations, and they may also help to understand and interpret observational systematics. We thus envision to carry out a full forward modelling of mock lightcone data, and a detailed comparison of the SZ signal of MTNG740 clusters on the lightcone with observations, in future work.

\section{Summary and outlook}
\label{sec:discussion}

In this introductory paper of the MillenniumTNG project, we have focused on the flagship full physics simulation that evolves a $500\,h^{-1}\mathrm{Mpc}$ (740 Mpc) cosmological box to $z=0$ with a baryonic mass resolution of $3.1\times 10^7\,\mathrm{M_\odot}$ and a physics model close to the IllustrisTNG model. In Section~\ref{sec:verification} we established that the MTNG740 hydrodynamical simulation is consistent with its IllustrisTNG predecessors as well as with recent observational data (see Figure~\ref{fig:tngmtng}). In particular, the color bimodality of galaxies is still present despite the slightly reduced resolution compared to TNG300. The transition from star-forming blue galaxies to quenched red galaxies occurs at a slightly higher stellar mass in MTNG740 than in TNG (see Figure~\ref{fig:galaxycolors}). The baryonic impact on the total matter matter power spectrum in MTNG740 is essentially identical to TNG100 as shown in Figure~\ref{fig:powerspectrum}.

We then presented a first analysis of the integrated properties of galaxy clusters in MTNG740. In Section~\ref{sec:scalingrelations} we discussed different galaxy cluster scaling relations in MTNG740 and compared them to the nearby galaxy cluster sample by \citet{Mulroy2019}. We found generally good agreement between MTNG740 and observed clusters. The normalisation of the richness of the observed galaxy clusters seems to be about a factor of two higher than the richness of the MTNG740 at fixed mass. However, an important caveat is that the mass estimates of the observed clusters are based on weak gravitational lensing mass estimates that might be biased.

We then looked at the internal structure of the MTNG740 clusters in Section~\ref{sec:clusterprofiles}. We examined internal profiles of hydrodynamical quantities as well as metalliticy and stellar mass of the MTNG740 galaxy clusters and compared them to observed profiles of nearby well-observed galaxy clusters from the X-COP survey \citep{Eckert2019, Ettori2019, Ghirardini2019}. We again found satisfactory agreement between the profiles of the MTNG740 clusters and observations. We found the main discrepancy in the inner part of the hydrodynamical profiles of clusters, that are able to reproduce non-cool core clusters, but not cool-core clusters, perhaps requiring a more physical AGN feedback model that reproduces both classes of clusters as observed. A possible refinement to the TNG galaxy model could be to include large-scale jets for the most massive systems.

In Section~\ref{sec:lightcones}, we considered the Compton-$y$ map of a deep lightcone up to redshift $z=5$. The average background value of the Compton-$y$ parameter we find is consistent with the latest constraints from Planck. We then compared the Compton-$y$ of a galaxy cluster at $z=0.25$ measured in different apertures. We showed that neither spherical nor cylindrical apertures fully describe the signal observed on the lightcone. In particular, for large apertures ($5\,R_{500}$) the Compton-$y$ background significantly contributes so that it cannot be ignored and requires matched filtering to derive an unbiased estimate of the cluster-intrinsic Compton-$y$ signal.

Having shown that the MTNG740 galaxy cluster population is overall consistent with observed galaxy clusters but shows some interesting discrepancies, we turned  in Section~\ref{sec:observations} to a comparison that stays fully in observational space and sidesteps the thorny issue of cluster mass estimates. We compared the galaxy clusters in MTNG740 with a set of galaxy clusters found in both, the Planck SZ cluster catalogue, and the RedMaPPer SDSS-8 catalogue. We compared richness and Compton-$y$, which are both observables that can also be directly extracted from the simulation. Overall, we found quite good agreement in Figure~\ref{fig:richness_sz}. Moreover, we confirmed in Figure~\ref{fig:clustergalaxies} that the population of cluster galaxies contributing to the richness estimate is consistent between simulation and observations. However, at the same time this has highlighted that at similar richness and Compton-$y$ parameter, the SZ-based $M_{500}$ mass estimates of the Planck cluster catalogue are $0.2\,{\rm dex}$ higher than the masses measured for the MTNG740 clusters. We discussed possibly explanations for this offset and remark that this discrepancy is completely consistent with the weak lensing and richness-based mass correction applied in more recent SZ cluster surveys \citep{Hilton2021}.

We conclude that MTNG740 offers a great opportunity to study and understand galaxies and galaxy clusters in their large-scale cosmological context. Also, the MTNG740 simulations can be used to interpret data from future large cosmological surveys, and to quantify and potentially correct for observational biases in cosmological measurements. We give further examples of such applications in our set of companion papers, and presently work on additional research in this direction in forthcoming studies.

\section*{Data availability}
The MillenniumTNG simulations will be made fully publicly available at \url{https://www.mtng-project.org/} in 2024. The data shown in the figures of this article will be shared upon reasonable request to the corresponding author. 

\section*{Acknowledgements}

We thank the anonymous referee for helpful comments that improved the paper. RP thanks Wilma Trick for help with the color scheme of the paper. The authors gratefully acknowledge the Gauss Centre for Supercomputing (GCS) for providing computing time on the GCS Supercomputer SuperMUC-NG at the Leibniz Supercomputing Centre (LRZ) in Garching, Germany, under project pn34mo. This work used the DiRAC@Durham facility managed by the Institute for Computational Cosmology on behalf of the STFC DiRAC HPC Facility, with equipment funded by BEIS capital funding via STFC capital grants ST/K00042X/1, ST/P002293/1, ST/R002371/1 and ST/S002502/1, Durham University and STFC operations grant ST/R000832/1. CH-A acknowledges support from the Excellence Cluster ORIGINS which is funded by the Deutsche Forschungsgemeinschaft (DFG, German Research Foundation) under Germany’s Excellence Strategy – EXC-2094 – 390783311. VS and LH acknowledge support by the Simons Collaboration on “Learning the Universe”. LH is supported by NSF grant AST-1815978. CP acknowledges support by the European Research Council under ERC-AdG grant PICOGAL-101019746. SB is supported by the UK Research and Innovation (UKRI) Future Leaders Fellowship [grant number MR/V023381/1].  



\bibliographystyle{mnras}


\bsp	
\label{lastpage}
\end{document}